\documentclass{article}

\usepackage[eandd, final]{neurips_2026}

\usepackage[utf8]{inputenc}
\usepackage[T1]{fontenc}
\usepackage{hyperref}
\usepackage{url}
\usepackage{booktabs}
\usepackage{amsfonts}
\usepackage{nicefrac}
\usepackage{microtype}
\usepackage{xcolor}

\usepackage{graphicx}
\usepackage{subcaption}
\usepackage{float}
\usepackage{amsmath}
\usepackage{amssymb}
\usepackage{mathtools}
\usepackage{amsthm}
\usepackage[capitalize,noabbrev]{cleveref}
\usepackage[textsize=tiny]{todonotes}

\theoremstyle{plain}

\theoremstyle{definition}

\theoremstyle{remark}

\title{ViroGym: Realistic Large-Scale Benchmarks for Evaluating Viral Proteins}

\author{%
  Yichen Zhou \\
  GSK plc; Technical University of Munich \\
  \texttt{yichen.x.zhou@gsk.com} \\
  \And
  Jonathan Golob \\
  GSK plc; University of Washington \\
  \And
  Amir Mohammad Karimi Mamaghan \\
  KTH Royal Institute of Technology \\
  \AND
  Stefan Bauer\thanks{Equal contribution.} \\
  Technical University of Munich; Helmholtz AI \\
  \And
  Patrick Schwab\footnotemark[1] \\
  GSK plc \\
}

\begin{document}

\maketitle

\begin{abstract}
Protein language models (pLMs) have shown strong potential for zero-shot prediction of missense variant effects, yet systematic benchmarking on viral proteins remains limited, a critical gap given the need for proactive tools that can anticipate emerging mutations ahead of experimental validation. Here we introduce ViroGym, a comprehensive benchmark evaluating pLMs across three tasks: 79 deep mutational scanning (DMS) assays covering eukaryotic viruses with 552,065 mutated sequences across 7 phenotypic readouts, 21 influenza neutralisation tasks, and a real-world pandemic prediction task for SARS-CoV-2. We benchmark well-established pLMs on fitness landscapes, antigenic diversity, and pandemic forecasting, and find that the ProGen2 family consistently achieves the strongest performance across all three tasks. Crucially, DMS and neutralisation performance reliably identifies models that generalise to real-world emergence, even though the mutation sets they surface barely overlap, revealing that complementary in vitro benchmarks capture the evolutionary constraints needed for real-world mutation forecasting. The data and code supporting this study are publicly available on GitHub at \url{https://github.com/GSK-AI/viroGym}
\end{abstract}

\section{Introduction}
The most clinically relevant respiratory viruses--such as influenza, SARS-CoV-2, and others--mutate at a rapid pace, challenging both the immune system and development of effective vaccines and treatments. Even with extensive near real-time genomic reporting systems, such as GISAID \cite{shu2017gisaid} and Nextstrain \cite{hadfield2018nextstrain}, public-health authorities must continually anticipate the future direction of these rapidly evolving pathogens, and mismatches between the predicted and actual trajectory carry real costs, from vaccines poorly matched to circulating strains to increased breakthrough infections, hospitalisations, and deaths.

A familiar example is the current vaccine development system for SARS-CoV-2 and influenza, which involves a semi-annual strain selection process recommended by the World Health Organization (WHO). This production system, especially for seasonal influenza vaccines, has remained largely unchanged for over 40 years \cite{wei2020next}. Moreover, the vaccine effectiveness (VE) of seasonal influenza vaccines, i.e.\ the observed reduction in medically attended, laboratory-confirmed illness among vaccinated relative to unvaccinated individuals as estimated by the CDC test-negative design \cite{centers2020cdc}, has ranged from 19\% to 60\% across the 2009 to 2025 seasons \cite{centers2020cdc}. The peak vaccine effectiveness for SARS-CoV-2 in autumn 2023 reached 50.6\% within 2--4 weeks but then declined to 13.6\%, largely due to the emergence of new variants \cite{kirsebom2024effectiveness}. Despite of suboptimal vaccine efficacy, manufacturers must produce and release vaccines within six months of WHO announcements.

Given the need to design, pilot, manufacture, and test vaccines against emerging strains, a proactive vaccine design framework is needed to enable scientists to initiate preparation for manufacturing prior to WHO strain announcements. The ideal framework should also be broad enough to cover viruses associated with infectious diseases, such as Zika virus, Hepatitis B virus, and Human Immunodeficiency Virus (HIV). With the demonstrated success of large language models (LLMs) in natural language and, increasingly, in modelling biological sequences \cite{hie2021learning, rives2021biological}, it is plausible that such a proactive framework could be effective.

LLMs trained to predict amino acid sequences, known as pLMs, have had success with estimating the functional impact and fitness consequences of candidate mutations without requiring prior evolutionary or epidemiological information \cite{meier2021language}, demonstrating its great potential in enabling early-stage anticipation of antigenic changes and supporting proactive vaccine design. While current pLMs have largely been validated on non-viral sequences, with most of the foundational model training explicitly masking viral sequences from training, testing, and validation sets. Therefore, there remains a gap in our understanding of how different pLMs perform with viral genomic sequences. A clear set of benchmarks relevant to modelling of viral evolution is a key step towards applying pLM to vaccine and antiviral development.

To address these limitations, we present ViroGym, a realistic large-scale benchmark designed to evaluate pLMs in zero-shot settings for global vaccine development. The benchmark consists of three core tasks:

\begin{samepage}
\begin{itemize}
\item \textbf{Mutational effect prediction}, which evaluates model ability to capture complex, non-linear correlations within viral genomic sequences and to infer the functional consequences of individual mutations.
\item \textbf{Antigenic diversity prediction}, which assesses model capacity to understand immune escape and strain differentiation.
\item \textbf{Pandemic prediction}, which identifies models with strong zero-shot generalization suitable for modelling mutations observed in natural viral evolution.
\end{itemize}
\end{samepage}

ViroGym includes over 552,065 mutated sequence readouts, 2,691 viral sequence--titer pairs, and 24,187 naturally occurring single-mutation frequency measurements. Of the DMS entries, 501,984 are single amino acid substitutions and 3,188 are deletions or indels (see Table~\ref{tab:mut_types} in Appendix~\ref{sec:appendix_dms}). It spans 13 virus types and 7 phenotypic categories, providing broad coverage across viral families and functional properties (see Table~\ref{detailed_types} in Appendix~\ref{sec:appendix_dms} for details). We note that SARS-CoV-2 comprises approximately 47\% of DMS assays (36 of 79) and Influenza A comprises approximately 22\%, reflecting the availability of large-scale experimental data for these pathogens. This composition is an inherent feature of the current published DMS landscape rather than a deliberate design choice. By providing clinical meaningful and rigorous benchmarks, ViroGym enables a more realistic assessment of model utility for vaccine and antiviral development.

\section{Related Work}
\textbf{ProteinGym.} ProteinGym is a benchmark suite for evaluating how well pLMs predict the functional effects of protein mutations. It aggregates large-scale DMS datasets across diverse proteins, mutation types, and functional assays, with biologically grounded evaluation metrics \cite{notin2023proteingym}. Most prediction tasks involve non-viral proteins: of 217 assays, 24 concern eukaryote-infecting viruses (predominantly influenza A, HIV, and SARS-CoV-2) and 7 concern bacteriophages. ViroGym includes all 24 eukaryote-infecting viral assays and broadens coverage to 13 virus types, adding under-represented families such as Nipah, Lassa, Rabies, Zika, Dengue, Hepatitis B, Hepatitis C, and enteroviruses.

\textbf{EVEREST.} EVEREST evaluates pLM performance on viral mutational fitness prediction using 45 curated viral DMS datasets. It reports that current models fail to produce reliable predictions for more than half of WHO-prioritised pandemic-threat viruses \cite{gurev2025variant}. Its focus on priority viruses means that many other available viral DMS assays are not included.

\textbf{DMS Correlation Studies.} Livesey and Marsh~\cite{livesey2025variant} collected 13 new DMS datasets from ProteinGym and evaluated 97 variant effect predictors (VEPs) across 36 human proteins. They observed a strong correspondence between DMS benchmark performance and the ability to classify clinical variants, particularly for predictors not trained on clinical data. These findings suggest that VEPs could complement, and in some cases partially substitute for, in vitro experiments in assessing variant effects.

\textbf{MSA-based evolutionary models.} A parallel line of work trains deep generative models on multiple sequence alignments (MSAs) rather than raw sequences. EVE~\cite{frazer2021disease} fits a variational autoencoder to a protein family's MSA to assign unsupervised fitness scores, while EVEscape~\cite{thadani2023learning} extends this framework to predict immune escape by combining evolutionary likelihood with structural accessibility and antibody epitope information. These models perform strongly on well-studied proteins, but their reliance on high-quality MSAs limits applicability to novel or under-sampled viral proteins. This is a practical bottleneck early in a pandemic, although increasing sequencing throughput may ease the constraint. ViroGym therefore focuses on single-sequence pLMs that require no MSA, while acknowledging MSA-based approaches as complementary tools.

\textbf{Structure- and indel-aware alternatives.} Two model classes lie outside our sequence-only scope and are complementary baselines: structure-based pipelines that score mutants from predicted structures (e.g.\ AlphaFold~\cite{jumper2021highly}) with molecular-dynamics or free-energy calculations, which are far costlier per variant and depend on structures that lag sequence availability; and indel-aware diffusion models such as DPLM~\cite{wang2024diffusion} and its evolution-aware variant DPLM-Evo, natural supplementary baselines on our indel-containing subset.

\textbf{DMS-guided viral forecasting.} Beyond benchmarking, recent work has explored whether DMS measurements can directly forecast real-world viral evolution. Dadonaite~et~al.~\cite{dadonaite2024spike} showed that Spike DMS fitness scores correlate with the subsequent global success of SARS-CoV-2 clades, linking in vitro readouts to lineage-level epidemiological outcomes. Kikawa~et~al.~\cite{kikawa_high-throughput_2025} demonstrated a similar correspondence for influenza, where high-throughput neutralisation measurements predict strain evolutionary success. ViroGym builds on these insights by testing whether pLMs can reproduce and extend this predictive signal without experimental data at inference time, and by systematically comparing their predictions with DMS assays and GISAID surveillance data.

\section{ViroGym}
The benchmark comprises 79 DMS assays, 21 sequencing-based neutralisation assays for influenza A, and a real-world prediction task derived from the Global Initiative on Sharing All Influenza Data (GISAID), which provides genomic surveillance data for SARS-CoV-2. A glossary of virology-specific terms (titer, antigenicity, indel, HA, RBD) is provided in Appendix~\ref{sec:glossary}.

Figure~\ref{virogym} illustrates the overall framework. We evaluate pLMs zero-shot by scoring amino acid sequences using suitable strategies, then rank them against experimental measurements (DMS and neutralisation) and naturally occurring GISAID mutations. This evaluation spans controlled in vitro experiments and real-world pandemic settings.

\begin{figure*}[ht]
  \vskip 0.2in
  \begin{center}
    \centerline{\includegraphics[width=\textwidth]{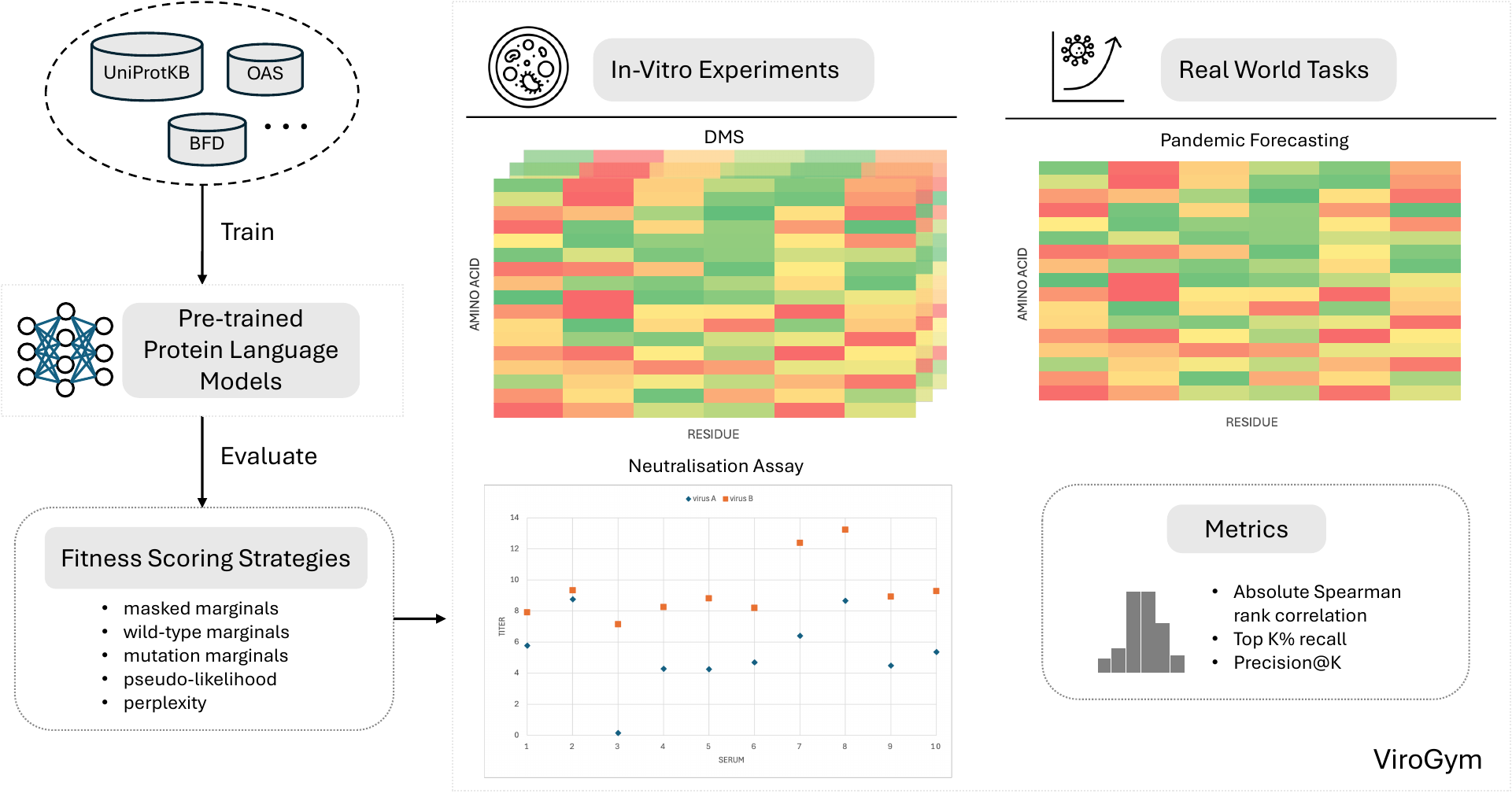}}
    \caption{
    ViroGym benchmark framework. The benchmark comprises in vitro experimental evaluation and real-world prediction tasks. The in vitro component uses DMS and neutralisation measurements to evaluate model predictions of protein functional effects. The real-world component evaluates SARS-CoV-2 pandemic forecasting using GISAID sequences, assessing generalisation from controlled wet-lab settings to natural viral evolution.
    }
    \label{virogym}
  \end{center}
\end{figure*}

\subsection{Dataset sources}
\textbf{DMS.} DMS is a high-throughput experimental technique that characterises a protein's functional landscape by systematically evaluating millions of single--amino-acid variants and mapping each mutation (genotype) to a measured functional property (phenotype) \cite{fowler_deep_2014}. ViroGym follows ProteinGym's assay selection guidelines and includes SARS-CoV-2, Influenza A, HIV, and 10 other viruses (Tables~\ref{DMS_all_reference} and \ref{DMS_all_reference_1}, Appendix~\ref{sec:appendix_dms}). Beyond ProteinGym's functional categories, ViroGym adds immune escape, a critical phenotype for viral proteins with direct relevance to vaccine and therapeutic development. 

\textbf{Neutralisation assay.} Traditional serological assays assess antibody neutralisation against a single viral strain per serum sample. In contrast, sequence-based high-throughput neutralisation assays quantify serum antibody responses through neutralisation titers across all relevant strains in a single experiment \cite{neu_loes2024high} (Table~\ref{neutralisation_all_reference}, Appendix~\ref{sec:appendix_neu}). These dense, sequence-resolved measurements allow machine learning models to jointly learn viral sequence variation and antigenic response. The resulting mappings between viral evolution and antibody-mediated immunity can help identify antigenically novel epitopes. All 21 neutralisation assays are newly curated.

\textbf{GISAID database.} GISAID is a global surveillance platform that monitors priority pathogens and facilitates the sharing of their genetic sequences and associated metadata \cite{shu2017gisaid}. This resource enables researchers to track viral evolution and transmission dynamics during epidemics and pandemics.

\subsection{Dataset processing}
\textbf{DMS.} Each dataset is processed independently into a common schema (mutation identifier, mutant sequence, DMS score) by remapping mutations to the reference sequence and removing missing values; scores are not globally normalised across assays. Where multiple measurements map to the same mutant (e.g.\ technical replicates, multiple sera) arithmetic means are taken. Influenza neutralisation titers are residualised against serum effects before export. Full processing and QC details are provided in Appendix~\ref{sec:appendix_processing}.

\textbf{Neutralisation assay.} We reviewed published information on assay participants to manually curate their vaccination histories. We then identified the vaccines each participant had received and obtained the corresponding HA1 sequences, enabling assessment of the antigenic coverage provided by these vaccines.

\textbf{GISAID database.} We collected all circulating SARS-CoV-2 sequences from GISAID spanning January 1, 2020, to December 31, 2025. We extracted all Spike protein mutations and recorded their observed occurrences. The resulting heat map shows mutation prevalence, including deletions, at each residue (Appendix~\ref{sec:heatmaps}).

\subsection{Baselines}
Just as language models learn grammar and contextual meaning from text, pLMs can learn biological rules and functional properties from amino acid sequences. The rapid growth of protein sequence data has enabled unsupervised training of representations that capture protein structure and biochemical properties, providing features for diverse biomedical applications~\cite{rives2021biological}.

We focus on single-sequence pLMs representative of current approaches: ESM-1 \cite{rives2019biological}, a first-generation pLM; ESM-1v \cite{meier2021language}, enabling zero-shot variant fitness prediction; ESM-2 \cite{lin2022language}, available in sizes from 8M to 15B parameters; and ProtT5 \cite{elnaggar2021prottrans}, an encoder-decoder architecture that captures amino acid context and provides embeddings for models such as VESPA and VESPAI \cite{marquet_embeddings_2022}. We also evaluate the ProGen2 suite \cite{nijkamp_progen2_2023}, which explores dataset and scale effects from antibody-specific models to the large BFD90-trained model; ProtGPT2 \cite{ferruz_protgpt2_2022}, designed for de novo protein generation; and Tranception \cite{notin2022tranception}, which robustly models protein fitness landscapes.

Our primary focus is accurate zero-shot prediction of variant fitness, as experimental characterisation of protein function can take weeks or months. During the COVID-19 pandemic, for example, structural analysis revealing atomic-level conformations of the SARS-CoV-2 receptor-binding domain (RBD) was completed one month after the first full genome sequences became available \cite{wrapp2020cryo}. To support timely early vaccine development while limiting data leakage, ViroGym focuses on single-sequence pLMs. We include only established versions with documented pre-training cutoffs prior to 2021, rather than the latest releases. This costs some accuracy: ESM3 \cite{hayes2025simulating}, for example, outperforms ESM2-650M on DMS under the same scoring strategy (wild-type marginals), but its training corpus extends past our evaluation window. Since most ViroGym data were collected after 2020, the pre-2021 cutoff is intended to measure genuine zero-shot generalisation rather than incidental exposure to assayed sequences during pre-training. Table~\ref{tab:leakage} in Appendix~\ref{sec:appendix_rankings} provides each model's training corpus, documented cutoff, viral-sequence retention, and leakage mitigation.

\subsection{Evaluation metrics}
We use two ranking-based metrics: \textbf{absolute Spearman rank correlation} to measure agreement between predicted and experimental rankings \cite{notin2023proteingym}, and \textbf{top-10\% recall} to focus evaluation on the highest-impact mutations (following ProteinGym convention). For the pandemic prediction task we additionally report \textbf{Precision@K}, measuring how many of the top-$K$ model predictions match the most prevalent real-world mutations. All reported numbers are macro-means over tasks, weighting each assay equally; equations for every metric and scoring strategy are given in Appendix~\ref{sec:appendix_metrics}.

\section{Results}
\subsection{Mutational effect prediction}
For encoder-based models we evaluate seven scoring strategies and find that semantic distance — the Euclidean distance between mean-pooled last-layer embeddings of wild-type and mutant sequences — consistently outperforms likelihood-based strategies (Table~\ref{ESM_stratgey}); decoder-only models (ProGen2, Tranception, ProtGPT2) use negative log-likelihood. Full details and ablations are in Appendix~\ref{sec:appendix_results}.

Table~\ref{DMS} presents aggregate mutational effect prediction results, with detailed per-model metrics in Appendix~\ref{sec:appendix_rankings}. ProGen2 achieves the strongest performance; Figure~\ref{DMS_PER_TASK} in Appendix~\ref{sec:appendix_figures} illustrates the per-task comparison with ESM2 15B. We assessed statistical significance using an exact two-sided paired sign test across all 79 shared DMS assays, with Benjamini--Hochberg correction applied separately for top-$k$ recall and absolute Spearman correlation (adjusted $p < 0.05$; Table~\ref{tab:dms_significance}, Appendix~\ref{sec:appendix_rankings}). ProGen2-XL ranks first on both aggregate metrics and significantly outperforms 21 of 24 competing models on recall. The three non-significant comparisons involve the closest ProGen2 variants (Base, BFD90) and Tranception L, where differences are small. Signed Spearman results preserve this ranking almost exactly ($\rho=0.997$; Appendix~\ref{sec:appendix_signed}), and a training-free BLOSUM62 baseline \cite{henikoff1992amino} also falls below ProGen2-XL (Table~\ref{DMS}).

\begin{table}[t]
  \caption{Performance of ESM2-650M under different scoring strategies. Results are reported as the average top 10\% recall and absolute Spearman's rank correlation between model predictions and experimental measurements, with standard deviation across 79 tasks.}
  \label{ESM_stratgey}
  \begin{center}
    \begin{small}
      \begin{sc}
        \begin{tabular}{lccccr}
          \toprule
          Strategy & Recall & Std. & Spearman & Std. \\
          \midrule
          masked   & 0.1145 & 0.0454 & 0.1089 & 0.1206 \\
          wildtype & 0.1123 & 0.0441 & 0.1063 & 0.1157 \\
          mutation & 0.1144 & 0.0432 & 0.1083 & 0.1180 \\
          grammar  & 0.1244 & 0.0452 & 0.1108 & 0.1170 \\
          semantic & \textbf{0.1382} & 0.0608 & \textbf{0.1678} & 0.1045 \\
          ratio    & 0.0809 & 0.0450 & 0.0912 & 0.0925 \\
          loss     & 0.1051 & 0.0557 & 0.1186 & 0.1253 \\
          \bottomrule
        \end{tabular}
      \end{sc}
    \end{small}
  \end{center}
\end{table}

\begin{table}[t]
  \caption{Zero-shot performance on the DMS benchmark. Results are reported as the average top 10\% recall and absolute Spearman's rank correlation between model scores and experimental measurements across all baselines, with standard deviation across 79 tasks.}
  \label{DMS}
  \begin{center}
    \begin{small}
      \begin{sc}
        \begin{tabular}{lcccr}
          \toprule
          Model & Recall & Std. & Spearman & Std. \\
          \midrule
          VESPAl     & 0.1681 & 0.1279 & 0.2561 & 0.1486 \\
          VESPA      & 0.1669 & 0.0971 & 0.2380 & 0.1351 \\
          Trancept.  & 0.1898 & 0.0857 & 0.2735 & 0.1482 \\
          ProtGPT2   & 0.1105 & 0.0382 & 0.1019 & 0.0768 \\
          ProGen2    & \textbf{0.1973} & 0.0913 & \textbf{0.2928} & 0.1595 \\
          ESM1v      & 0.1461 & 0.0667 & 0.1950 & 0.1018 \\
          ESM1       & 0.1466 & 0.0590 & 0.1978 & 0.1030 \\
          ESM2       & 0.1389 & 0.0719 & 0.1730 & 0.1110 \\
          \midrule
          BLOSUM62   & 0.1580 & --     & 0.2130 & --     \\
          \bottomrule
        \end{tabular}
      \end{sc}
    \end{small}
  \end{center}
\end{table}

\subsection{Antigenic diversity prediction}
Influenza vaccine strains are selected according to how far circulating viruses have drifted from previously dominant strains, integrating both genetic and antigenic evolution \cite{smith2004mapping, fouchier2010use}. Modelling genetic evolution from historical data is routine, typically through phylogenetic trees constructed using maximum likelihood estimation (MLE) \cite{felsenstein1973maximum}, but capturing antigenic evolution still requires wet-lab measurements. If a pLM can predict whether an emerging strain is likely to be covered by a given vaccine strain, it could accelerate the vaccine development cycle.

To evaluate this capability, we use 21 influenza neutralisation assays to assess whether pLMs detect antigenic differences among viral strains. These assays generally use haemagglutination inhibition, in which antibody titers serve as a proxy for antigenic similarity; for example, a titer of 1:40 is often considered indicative of adequate immune coverage \cite{hannoun2004immunogenicity}. Within this framework, we query pLMs to estimate antigenic similarity between vaccine strains and newly isolated strains. Higher similarity scores should, conceptually, correspond to stronger expected vaccine-mediated protection.

We assess performance using two sequence-based readouts, motivated by the hypothesis that pLMs place biologically similar HA1 sequences close together in their learned representation or distribution. These sequence-based quantities may therefore serve as proxies for antigenic novelty. For encoder-only models (ESM, ProtT5, VESPA), we use the Euclidean distance between mean-pooled last-layer embeddings of vaccine-strain HA1 ($v$) and circulating-strain HA1 ($s$), $d(v,s)=\lVert \bar{h}(v)-\bar{h}(s)\rVert_2$, as predicted antigenic distance. For decoder-only models (ProGen2, Tranception, ProtGPT2), there is no clean pairwise sequence embedding, so we instead use the circulating strain's length-normalised perplexity,
\[
  \mathrm{PPL}(s)=\exp\!\Big(-\tfrac{1}{|s|}\textstyle\sum_i \log p_\theta(s_i\mid s_{<i})\Big),
\]
as a proxy rather than a calibrated pairwise distance to the vaccine strain. Sequences the model finds unexpected relative to its training distribution are hypothesised to be more antigenically novel. We correlate predictions with experimental HI log$_2$ titre to test whether pLMs approximate fine-grained antigenic relationships that could, in principle, guide vaccine strain selection.

Performance differences among models are marginal, with Tranception M nominally leading Table~\ref{Neutralisation}. Appendix~\ref{sec:appendix_figures} provides task-wise results, including a detailed breakdown for Tranception M in Figure~\ref{TM_AN}. Exact two-sided paired sign tests across all 21 tasks, with Benjamini--Hochberg correction, find 0 of 21 comparisons significant against either Tranception M or ProGen2-XL ($p_{\mathrm{adj}}\ge0.076$). We therefore do not claim a statistically significant winner: at this scale, the neutralisation panel is best viewed as a difficult diagnostic benchmark rather than a leaderboard, with substantial room for improvement.

\begin{table}[t]
  \caption{Zero-shot neutralisation prediction results. Reported values are the average absolute Spearman's rank correlation between model predictions and experimental measurements across all baseline methods.}
  \label{Neutralisation}
  \begin{center}
    \begin{small}
      \begin{sc}
        \begin{tabular}{lcccr}
          \toprule
          Model Name  & Spearman & Std. \\
          \midrule
          ProtT5      & 0.1961 & 0.206  \\
          Tranception & \textbf{0.2316} & 0.1696 \\
          ProtGPT2    & 0.2018 & 0.1845 \\
          ProGen2     & 0.2250 & 0.1852 \\
          ESM1v       & 0.2282 & 0.2043 \\
          ESM1        & 0.2222 & 0.2098 \\
          ESM2        & 0.2267 & 0.1840 \\
          \midrule
          BLOSUM62    & 0.2244 & 0.1571 \\
          \bottomrule
        \end{tabular}
      \end{sc}
    \end{small}
  \end{center}
\end{table}

\subsection{Pandemic prediction}
\label{sec:pandemic}
pLMs are increasingly viewed as general-purpose tools for protein prediction, potentially replacing traditional multiple sequence alignment (MSA) methods \cite{weissenow2025protein}. Their capabilities include generating representations for secondary and tertiary structure prediction and inferring biochemical properties without labelled data \cite{rives2021biological}, identifying conserved residues without MSAs \cite{marquet_embeddings_2022}, and predicting missense mutation effects \cite{meier2021language}. Many perform strongly on in vitro benchmarks, but a critical question remains: can they generalise effectively to in vivo or real-world environments?

To connect in vitro results to real-world outcomes, we test whether pLMs can identify dominant circulating mutations using only the target SARS-CoV-2 Spike protein sequence. We use marginal single-mutation frequency relative to Wuhan-Hu-1, rather than lineage growth, as the prediction target for three reasons. First, mutations reaching high marginal frequency are enriched for positive selection: independent acquisition of the same substitution across phylogenetically distinct lineages, known as \emph{convergent evolution}, provides strong evidence of a genuine fitness advantage rather than passive genetic hitchhiking on a successful background~\cite{korber2020d614g, martin2021n501y, kistler2022convergent}. Second, lineage forecasting is complicated by the non-linear phylodynamics of SARS-CoV-2. Successful variants do not always descend from the currently dominant strain~\cite{markov2023evolution, carabelli2023variant}: Omicron BA.1, for example, diverged from a basal lineage predating Delta rather than descending from it. Substantial epistatic differences across these backgrounds make reference-sequence selection and epistatic control difficult. Third, individual substitutions are more tractable and actionable targets than lineages. Maher et al.\ demonstrate that single amino-acid changes destined to spread can be identified up to four months before reaching high frequency~\cite{maher2022predicting}, while Thadani et al.\ show that mutation-level escape forecasts trained on pre-pandemic sequence data alone are competitive with high-throughput experimental scans across multiple viruses~\cite{thadani2023evescape}. This lead time is directly relevant to vaccine antigen selection and therapeutic target prioritisation, where decisions must precede variant dominance by several months.

Using Wuhan-Hu-1 as a consistent reference across all periods allows us to evaluate whether pLMs identify mutations that proved successful in circulation, independently of their lineage of origin. Appendix~\ref{sec:heatmaps} provides heat maps of predicted in silico fitness scores for each amino acid at each residue across all baselines. ProGen2-XL achieves the strongest performance across all three pandemic metrics (Table~\ref{GISAID}), consistent with its dominance on DMS (Table~\ref{DMS}) and competitive neutralisation results (Table~\ref{Neutralisation}). Tranception is consistently the second-best family across all three tasks, leading neutralisation and remaining statistically indistinguishable from ProGen2-XL on DMS recall.
\begin{table}[t]
  \caption{Zero-shot pandemic prediction results. Metrics reported for all baselines include Top 10\% Recall, absolute Spearman's rank correlation, and Precision@3 between model scores and mutation frequencies from GISAID. Evaluated on 24,187 substitution mutations (deletion mutations excluded for fair cross-model comparison, as VESPA/VESPAl cannot score deletions; results on all 25,460 mutations excluding VESPA/VESPAl are in Appendix Table~\ref{gisaid_all_models}). For multi-size families the best-performing variant is shown. BLOSUM62 and Conservation (MSA, from a coronavirus Spike homolog alignment \cite{notin2023proteingym}) are training-free baselines.}
  \label{GISAID}
  \begin{center}
    \begin{small}
      \begin{sc}
        \begin{tabular}{lccc}
          \toprule
          Model Name       & Recall & Spearman & Precision@3 \\
          \midrule
          VESPAl           & 0.2953 & 0.3158 & 0.00 \\
          VESPA            & 0.3238 & 0.3616 & 0.00 \\
          Tranception L    & 0.3743 & 0.4014 & 0.33 \\
          ProtGPT2         & 0.0968 & 0.0495 & 0.00 \\
          ProGen2 XL       & \textbf{0.4347} & \textbf{0.4018} & \textbf{0.33} \\
          ESM1v            & 0.2676 & 0.2682 & 0.00 \\
          ESM1 (85M)       & 0.2630 & 0.2500 & 0.00 \\
          ESM2 (35M)       & 0.2667 & 0.2580 & 0.00 \\
          \midrule
          BLOSUM62           & 0.3130 & 0.3840 & 0.00 \\
          Conservation (MSA) & 0.2063 & 0.3616 & 0.33 \\
          \bottomrule
        \end{tabular}
      \end{sc}
    \end{small}
  \end{center}
\end{table}

\begin{figure}[h!]
  \begin{center}
    \includegraphics[width=\textwidth]{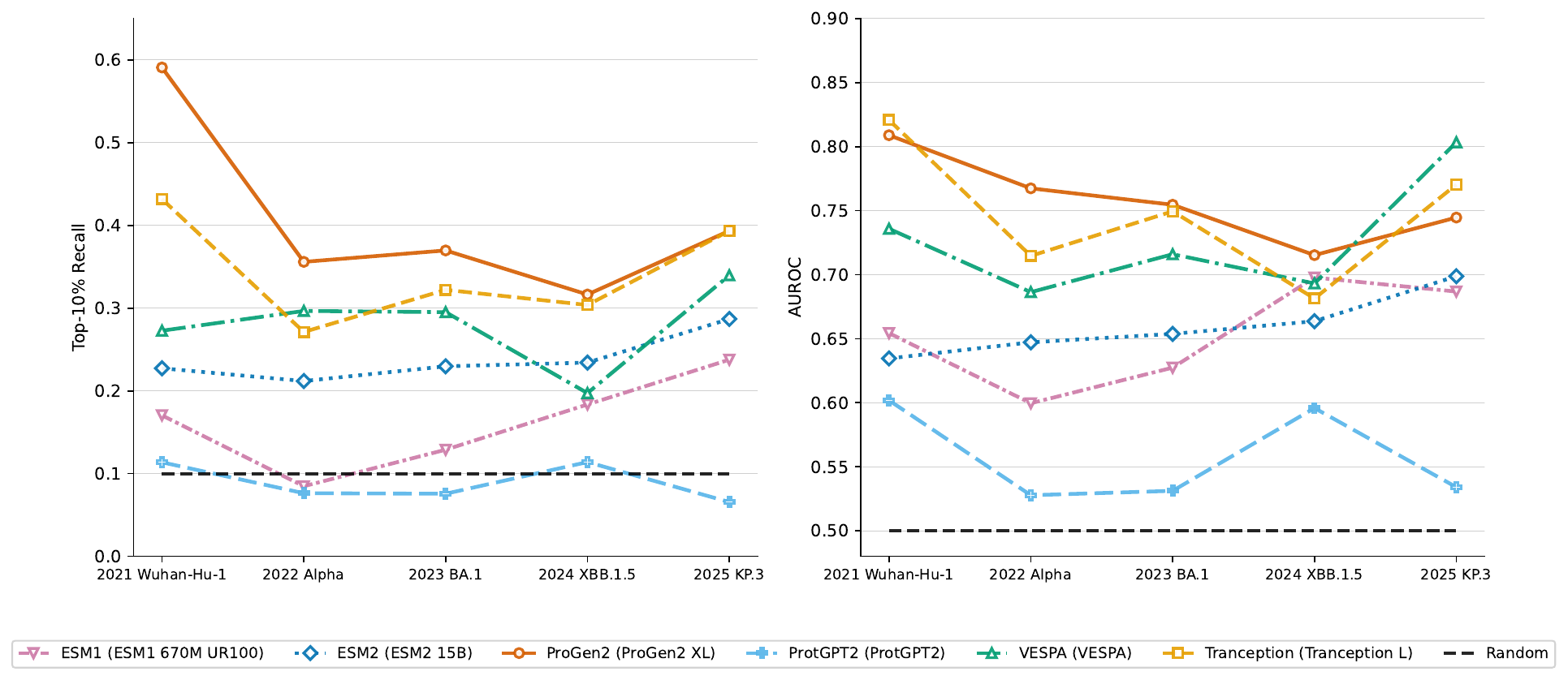}
    \caption{Pseudo-temporal validation across five SARS-CoV-2 variant eras. Each panel shows the metric for the best-performing variant from each model family as a function of the reference era. Deletion mutations are excluded for fair cross-model comparison. The dashed black line indicates the expected performance of a random predictor.}
    \label{fig:temporal}
  \end{center}
\end{figure}

These aggregate rankings use a static reference and pool mutations across the full pandemic. A complementary question is whether the ordering holds as the dominant strain shifts from Wuhan-Hu-1 through successive variant eras. We assess this through pseudo-temporal validation across five eras (2021 Wuhan-Hu-1, 2022 Alpha, 2023 BA.1, 2024 XBB.1.5, 2025 KP.3). For each era, models score all single-amino-acid substitutions relative to that period's dominant reference sequence. The binary ground truth, \textit{is\_emergent}, labels mutations that reached high global frequency in the subsequent window (defined in \ref{sec:appendix_processing}). Figure~\ref{fig:temporal} shows Top-10\% Recall and AUROC for each model family's best-performing variant across all five windows.

The result shows ProGen2 and Tranception consistently rank above ESM-family and ProtGPT2 models across all time windows. All top models maintain AUROC well above the random baseline (0.5) through the 2025 KP.3 era, demonstrating that the predictive signal does not degrade sharply as the reference sequence drifts from Wuhan-Hu-1. ESM-family and ProtGPT2 models trail consistently, with mean recalls below 0.25, suggesting that models pre-trained on large and diverse protein databases (ProGen2, Tranception) are better calibrated to emergent mutation frequency than those trained on curated or narrower sequence sets.

We next investigate how pLM predictions relate to DMS assays and real-world circulating mutations, restricting the analysis to the RBD. Among the top-10 RBD mutations, ESM2-650M has no overlap with GISAID or DMS, while ProGen2-XL shares $\sim$40\% with the most prevalent GISAID mutations and $\sim$20\% with DMS top hits; DMS and GISAID share only $\sim$10\% (Figure~\ref{LINK}, Appendix~\ref{sec:appendix_figures}). This low DMS--GISAID overlap reflects the mechanistic gap between a single controlled experimental condition and the composite in vivo fitness on which natural selection acts. The $\sim$20\% pLM--DMS intersection includes N501Y, a key driver of Alpha variant transmissibility through enhanced ACE2 binding~\cite{liu2022n501y}. Substitutions supported by both signals are therefore particularly high-confidence candidates for surveillance and vaccine antigen prioritisation.

\begin{figure}[h!]
  \centering
  \includegraphics[width=0.65\linewidth]{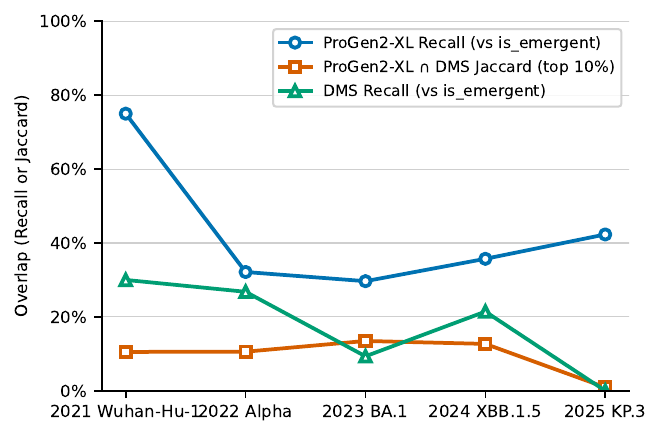}
  \caption{Temporal overlap of top-10\% RBD mutations (Spike positions 331--531) across five SARS-CoV-2 variant eras. \textbf{Blue:} Top-10\% Recall of ProGen2-XL predictions against \textit{is\_emergent} ground truth. \textbf{Red:} Top-10\% Recall of era-matched DMS binding assays against the same ground truth. \textbf{Green:} Jaccard similarity between the ProGen2-XL and DMS top-10\% prediction sets.}
  \label{fig:temporal_overlap}
\end{figure}

To examine how this relationship evolves across variant eras, we compute the top-10\% Recall of era-matched DMS binding assays and ProGen2-XL against \textit{is\_emergent}, alongside the Jaccard similarity between their top-10\% prediction sets (Figure~\ref{fig:temporal_overlap}). ProGen2-XL maintains recall between 30\% and 75\% throughout the pandemic, while DMS recall falls from 30\% in 2021 to 0\% by 2025 KP.3. This is consistent with DMS measuring fitness under fixed experimental conditions that become less representative as the virus diverges. The pLM--DMS Jaccard similarity remains low ($\sim$11\%) across 2021--2024, confirming that the two sources identify largely non-overlapping RBD candidates and provide complementary signals.

\section{Discussion}
ViroGym introduces an evaluation framework for viral proteins spanning mutational effect prediction, antigenic diversity prediction, and pandemic prediction, with the goal of linking in vitro experiments to real-world outcomes. DMS datasets provide a detailed view of protein fitness landscapes under controlled conditions, but real-world protein evolution, particularly in viruses, is shaped by additional constraints. For example, immune imprinting from early antigen exposure can bias antibody responses towards conserved epitopes, influencing vaccine strain selection and subsequently shaping viral evolutionary trajectories.

Our analysis highlights three considerations for improving pLMs on viral proteins. First, handling insertions and deletions (indels), which often disrupt protein function, remains challenging. To our knowledge, only ESM models explicitly encode deletions as vocabulary tokens, and excluding deletion-containing sequences yields modest gains across models (Table~\ref{no_del_tasks}, Appendix~\ref{sec:appendix_rankings}). Second, unlike sequence-based pLMs trained solely on UniProtKB, ViroGym's best-performing model, ProGen2, may benefit from joint pre-training on UniProtKB and BFD. Third, viral proteins are often longer than typical proteins, whereas most pLMs have context windows of fewer than 1024 residues. These observations suggest that expanded token representations, larger and more diverse training datasets, and longer contexts could improve performance by better capturing viral protein variation and long-range dependencies.

Our central finding is that in vitro and in silico signals are complementary, not redundant: DMS and neutralisation performance predicts which models generalise to real-world emergence, even though the mutations they surface barely overlap (Table~\ref{tab:dms_pandemic_corr}, Appendix~\ref{sec:appendix_results}). Low overlap among DMS, neutralisation, pLM, and GISAID signals is therefore not a limitation but a feature: each captures facets of viral fitness that the others miss. Combining fitness, antigenic, and evolutionary signals provides a principled strategy for mutation prioritisation, antigen selection, and early warning of emerging variants.

\section{Limitations and Future Work}
A limitation is that pandemic prediction and its real-world validation are specific to SARS-CoV-2 Spike mutations. Reliable, large-scale mutation frequency data for Influenza A and other viruses are substantially harder to obtain from GISAID, making cross-viral extension infeasible without dedicated surveillance-data curation. The finding that DMS-selected pLMs transfer well to real-world mutation prediction should therefore be interpreted as SARS-CoV-2-specific. Whether DMS--pLM complementarity generalises to other viral families remains an important direction for future work. Similarly, pseudo-temporal validation across SARS-CoV-2 variant eras provides encouraging evidence of temporal generalisation, but lineage-level temporal forecasting must account for epistatic context and lineage emergence dynamics and remains a complex open challenge.

Two further caveats apply. First, the pandemic-prediction ground truth \textit{is\_emergent} is a per-mutation marginal frequency, conflating a mutation's direct fitness contribution with epistatic hitchhiking on successful backgrounds; our convergent-evolution rationale (Section~\ref{sec:pandemic}) only partly mitigates this. Second, although within-study aggregation and per-assay rank metrics remove cross-laboratory scale differences, residual platform-specific noise-ceiling bias persists in the aggregated DMS data. ViroGym scores are therefore intended to prioritise candidates for wet-lab follow-up, not to replace experimental or epidemiological validation.

No model approaches saturation on any task, though future releases should still add post-2025 surveillance windows and further viruses. ViroGym also reports point estimates rather than calibrated confidences. Uncertainty signals that require no retraining, such as masked-marginal entropy, variance across model families, and calibration of top-$k$ lists against held-out eras, may indicate which predictions merit wet-lab follow-up. We leave their evaluation to future work.

\section*{Acknowledgement}
Computing resources for the ESM3 evaluation were provided as follows: The authors thank the Gauss Centre for Supercomputing e.V. (GCS) for providing computing time on the Supercomputer JUPITER at the Jülich Supercomputing Centre via the John von Neumann Institute for Computing. JUPITER is supported by the EuroHPC JU and GCS through funding by the European Commission, the German Federal Ministry of Research, Technology and Space, and the Ministry of Culture and Science of the State of North Rhine-Westphalia.

\bibliography{citation}
\bibliographystyle{abbrvnat}

\clearpage
\appendix

\section{Glossary}
\label{sec:glossary}
For readers outside virology we briefly define recurring terms. \emph{Titer}: the reciprocal of the highest serum dilution that still inhibits the virus in an assay, used as a proxy for antibody potency. \emph{Antigenicity}: the degree to which a viral protein is recognised by host antibodies; two strains are antigenically distant if antibodies to one neutralise the other poorly. \emph{Indel}: an insertion or deletion of one or more residues, as opposed to a substitution. \emph{HA}: haemagglutinin, the principal surface antigen of influenza A. \emph{RBD}: the receptor-binding domain of the SARS-CoV-2 Spike protein (residues 331--531), the primary target of neutralising antibodies.

\section{Evaluation Metrics and Scoring Strategies}
\label{sec:appendix_metrics}

This section provides definitions for every metric reported in the main text and the seven scoring strategies compared in Table~\ref{ESM_stratgey}.

\subsection{Metrics}
\label{sec:appendix_metric_defs}

Consider a single task with $n$ scored variants, experimental ground-truth values $y\in\mathbb{R}^n$ and model scores $\hat{y}\in\mathbb{R}^n$. All scores are oriented so that a larger value means higher predicted fitness (DMS), higher predicted antigenic novelty (neutralisation), or higher predicted real-world frequency (pandemic); the per-family sign conventions are listed in Table~\ref{tab:sign_convention}.

\paragraph{Absolute Spearman rank correlation.} With $\mathrm{rk}(\cdot)$ denoting the rank transform,
\[
  |\rho_s| \;=\; \left|\frac{\operatorname{cov}\!\big(\mathrm{rk}(\hat{y}),\,\mathrm{rk}(y)\big)}{\sigma_{\mathrm{rk}(\hat{y})}\,\sigma_{\mathrm{rk}(y)}}\right|.
\]
We follow the absolute-value convention of \citet{meier2021language}, which accommodates the mixed directionality of DMS readouts and scoring strategies; signed values for all 25 baselines are reported in Appendix~\ref{sec:appendix_signed}.

\paragraph{Top-10\% Recall.} Let $\mathrm{top}_k(v)$ be the index set of the $k$ largest entries of a vector $v$, and let $k=\lceil 0.1n\rceil$. Then
\[
  \mathrm{Recall@10\%} \;=\; \frac{\big|\mathrm{top}_k(\hat{y}) \cap \mathrm{top}_k(y)\big|}{k},
\]
i.e.\ the fraction of the experimentally top-10\% variants that the model also ranks in its top 10\%.

\paragraph{Precision@K.} For the pandemic-prediction task we additionally report, with $K=3$,
\[
  \mathrm{Precision@}K \;=\; \frac{\big|\mathrm{top}_K(\hat{y}) \cap \mathrm{top}_K(y)\big|}{K},
\]
where $y$ holds the observed GISAID mutation frequencies.

\paragraph{Binary (pseudo-temporal) variant.} In the five-era evaluation the ground truth is the binary label $z\in\{0,1\}^n$ (\texttt{is\_emergent}; Appendix~\ref{sec:appendix_processing}) rather than a continuous frequency. There we use
\[
  \mathrm{Recall@10\%} = \frac{\sum_{i\in \mathrm{top}_k(\hat{y})} z_i}{\sum_i z_i}, \qquad
  \mathrm{Precision@}K = \frac{\sum_{i\in \mathrm{top}_K(\hat{y})} z_i}{K}, \quad K=\textstyle\sum_i z_i,
\]
with $k=\lfloor 0.1n\rfloor$, alongside the standard AUROC of $\hat{y}$ against $z$. A random predictor attains $\mathrm{Recall@10\%}=0.10$ and $\mathrm{AUROC}=0.5$, which are the dashed baselines in Figure~\ref{fig:temporal}.

\paragraph{Aggregation.} Unless stated otherwise, a reported number is the \emph{macro-mean} of the per-task metric over the $T$ tasks of a benchmark ($T=79$ for DMS, $T=21$ for neutralisation, $T=5$ for the pseudo-temporal eras), $\frac{1}{T}\sum_{t=1}^{T} M_t$, with each task weighted equally regardless of its variant count; quoted standard deviations are taken across tasks. A variant-count-weighted alternative is given in Table~\ref{tab:dms_weighted}.

\subsection{Scoring strategies}
\label{sec:appendix_scoring}

Let $\theta$ denote a pLM, $x^{\mathrm{wt}}$ the wild-type sequence, and consider the substitution of residue $a$ at position $p$ by residue $b$, giving the mutant sequence $x^{\mathrm{mt}}$. We write $x_{\setminus p}$ for the sequence with position $p$ masked and $\bar{h}(\cdot)$ for the mean-pooled last-layer embedding. The first four strategies follow \citet{meier2021language}.

\begin{enumerate}
  \item \textbf{Masked marginal} --- position $p$ is masked and the mutant is scored against the wild-type residue under the resulting distribution:
  \[
    s_{\mathrm{masked}} = \log p_\theta\big(x_p = b \mid x^{\mathrm{wt}}_{\setminus p}\big) - \log p_\theta\big(x_p = a \mid x^{\mathrm{wt}}_{\setminus p}\big).
  \]
  \item \textbf{Wild-type marginal} --- the same log-odds, but conditioned on the unmasked wild-type sequence, requiring a single forward pass per protein:
  \[
    s_{\mathrm{wt}} = \log p_\theta\big(x_p = b \mid x^{\mathrm{wt}}\big) - \log p_\theta\big(x_p = a \mid x^{\mathrm{wt}}\big).
  \]
  \item \textbf{Mutant marginal} --- the same log-odds conditioned on the unmasked mutant sequence, $s_{\mathrm{mt}} = \log p_\theta(x_p = b \mid x^{\mathrm{mt}}) - \log p_\theta(x_p = a \mid x^{\mathrm{mt}})$.
  \item \textbf{Grammaticality} --- the mutant log-probability from the masked pass taken on its own, without reference to the wild-type residue, $s_{\mathrm{gram}} = \log p_\theta(x_p = b \mid x^{\mathrm{wt}}_{\setminus p})$.
  \item \textbf{Semantic distance} --- the embedding displacement induced by the mutation, negated so that larger means more fit:
  \[
    s_{\mathrm{sem}} = -\big\lVert \bar{h}(x^{\mathrm{mt}}) - \bar{h}(x^{\mathrm{wt}}) \big\rVert_2 .
  \]
  \item \textbf{Ratio} --- the probability ratio at the mutated position under the wild-type pass, $s_{\mathrm{ratio}} = p_\theta(x_p = b \mid x^{\mathrm{wt}}) \,/\, p_\theta(x_p = a \mid x^{\mathrm{wt}})$.
  \item \textbf{Loss} --- the negative mean cross-entropy of the mutant sequence under the model, $s_{\mathrm{loss}} = -\frac{1}{|x^{\mathrm{mt}}|}\sum_i -\log p_\theta\big(x^{\mathrm{mt}}_i \mid x^{\mathrm{mt}}_{\setminus i}\big)$, so that sequences the model finds more probable score higher.
\end{enumerate}

Decoder-only models (ProGen2, Tranception, ProtGPT2) expose no masked conditional or sequence embedding, and are scored by the length-normalised log-likelihood of the mutant sequence, $\frac{1}{|x^{\mathrm{mt}}|}\sum_i \log p_\theta(x^{\mathrm{mt}}_i \mid x^{\mathrm{mt}}_{<i})$. For the neutralisation task the corresponding quantities are defined on whole HA1 sequences rather than on single substitutions (Table~\ref{Neutralisation}).

\section{Benchmark Construction Details}
\label{sec:appendix_construction}

This section provides dataset provenance and phenotype breakdowns for all three ViroGym tasks. Tables~\ref{DMS_all_reference} and~\ref{DMS_all_reference_1} list all 79 DMS assays with their virus, phenotype, sequence count, and source reference; an asterisk (*) marks overlap with ProteinGym and a dagger ($\dagger$) marks overlap with EVEREST. Table~\ref{detailed_types} summarises the distribution across virus families and phenotypic categories. Table~\ref{neutralisation_all_reference} lists the 21 influenza~A neutralisation assays, while Table~\ref{non_indel_types} classifies the 79 DMS tasks by indel status for the deletion-sensitivity analysis in the main text.

\subsection{DMS Datasets}
\label{sec:appendix_dms}

\begin{table}[H]
  \caption{Sources of DMS benchmark datasets (part 1 of 2). Mutational effect prediction tasks are based on DMS assays. Datasets marked * overlap with ProteinGym; datasets marked $\dagger$ overlap with EVEREST.}
  \label{DMS_all_reference}
  \centering
    \begin{tabular}{lccr c}
      \toprule
      Dataset & Phenotype & Sequences & Reference \\
      \midrule
      *ZIKV & viral growth & 10{,}080 & \cite{sourisseau2019deep} \\
      ZIKV & immune escape & 9{,}576 & \cite{sourisseau2019deep}, \cite{kikawa2023effect} \\
      RABV & immune escape & 7{,}220 & \cite{aditham2025deep} \\
      RABV & cell entry & 8{,}614 & \cite{aditham2025deep} \\
      $^{\dagger}$NIPAH& binding & 8{,}654 & \cite{larsen2025functional} \\
      $^{\dagger}$NIPAH& immune escape & 8{,}690 & \cite{larsen2025functional} \\
      $^{\dagger}$NIPAH& cell entry & 10{,}090 & \cite{larsen2025functional} \\
      $^{\dagger}$LASV& immune escape & 4{,}105 & \cite{carr2024deep} \\
      $^{\dagger}$LASV& cell entry & 8{,}649 & \cite{carr2024deep} \\
      HIV B520 & immune escape & 7{,}720 & \cite{radford2023mapping} \\
      HIV B520 & cell entry & 7{,}038 & \cite{radford2025comprehensive} \\
      HIV B520 & immune escape & 7{,}033 & \cite{radford2023mapping}, \cite{radford2025comprehensive} \\
      HIV TRO11 & cell entry & 12{,}268 & \cite{radford2025comprehensive} \\
      HIV TRO11 & immune escape & 7{,}769 & \cite{radford2025comprehensive} \\
      *HIV HXB2 & viral growth & 2{,}147 & \cite{fernandes2016functional} \\
      *HIV BRU/LAI & viral growth & 1{,}577 & \cite{fernandes2016functional} \\
      *$^{\dagger}$HIV strain896 & viral growth & 375 & \cite{duenas2016saturation} \\
      *HIV BRU/LAI & viral growth & 12{,}863 & \cite{haddox2016experimental} \\
      *$^{\dagger}$HIV & viral growth & 12{,}729 & \cite{haddox2018mapping} \\
      *$^{\dagger}$HIV B520 & viral growth & 12{,}577 & \cite{haddox2018mapping} \\
      HBV & fitness & 17{,}692 & \cite{yu2024deep} \\
      *$^{\dagger}$SCV2 RBD Wuhan hu & binding & 3{,}978 & \cite{starr2020deep} \\
      *$^{\dagger}$SCV2 RBD Wuhan hu & expression & 3{,}973 & \cite{starr2020deep} \\
      SCV2 RBD Alpha & binding & 3{,}937 & \cite{starr2022shifting} \\
      SCV2 RBD Alpha & expression & 3{,}931 & \cite{starr2022shifting} \\
      SCV2 RBD Beta & binding & 4{,}005 & \cite{starr2022shifting} \\
      SCV2 RBD Beta & expression & 4{,}003 & \cite{starr2022shifting} \\
      SCV2 RBD Delta & binding & 4{,}019 & \cite{starr2022shifting} \\
      SCV2 RBD Delta & expression & 4{,}019 & \cite{starr2022shifting} \\
      SCV2 RBD Eta & binding & 3{,}944 & \cite{starr2022shifting} \\
      SCV2 RBD Eta & expression & 3{,}947 & \cite{starr2022shifting} \\
      SCV2 RBD Omicron BA.1 & binding & 4{,}020 & \cite{starr2022deep} \\
      SCV2 RBD Omicron BA.1 & expression & 4{,}020 & \cite{starr2022deep} \\
      \bottomrule
      \multicolumn{4}{l}{* from ProteinGym; $\dagger$ from EVEREST}
    \end{tabular}
\end{table}

\begin{table}[H]
  \caption{Sources of DMS benchmark datasets (part 2 of 2). Datasets marked * overlap with ProteinGym; datasets marked $\dagger$ overlap with EVEREST.}
  \label{DMS_all_reference_1}
  \centering
    \begin{tabular}{lccr c}
      \toprule
      Dataset & Phenotype & Sequences & Reference \\
      \midrule
      SCV2 RBD Omicron BA.2 & binding & 4{,}002 & \cite{starr2022deep} \\
      SCV2 RBD Omicron BA.2 & expression & 4{,}002 & \cite{starr2022deep} \\
      SCV2 RBD Omicron BQ.1.1 & binding & 4{,}203 & \cite{taylor2023deep} \\
      SCV2 RBD Omicron BQ.1.1 & expression & 4{,}202 & \cite{taylor2023deep} \\
      $^{\dagger}$SCV2 RBD Omicron XBB.1.5& binding & 4{,}212 & \cite{taylor2023deep} \\
      $^{\dagger}$SCV2 RBD Omicron XBB.1.5& expression & 4{,}212 & \cite{taylor2023deep} \\
      SCV2 RBD Omicron BA.2.86 & binding & 4{,}196 & \cite{taylor2024deep} \\
      SCV2 RBD Omicron BA.2.86 & expression & 4{,}196 & \cite{taylor2024deep} \\
      SCV2 RBD Omicron EG.5 & binding & 4{,}213 & \cite{taylor2024deep} \\
      SCV2 RBD Omicron EG.5 & expression & 4{,}213 & \cite{taylor2024deep} \\
      SCV2 RBD Omicron FLip & binding & 4{,}215 & \cite{taylor2024deep} \\
      SCV2 RBD Omicron FLip & expression & 4{,}215 & \cite{taylor2024deep} \\
      SCV2 Wuhan hu & immune escape & 2{,}123 & \cite{cao2022omicron} \\
      SCV2 RBD Omicron XBB.1.5 & immune escape & 3{,}032 & \cite{dadonaite2025sars} \\
      SCV2 RBD Omicron XBB.1.5 & cell entry & 3{,}918 & \cite{dadonaite2025sars} \\
      SCV2 Omicron XBB.1.5& immune escape & 6{,}931 & \cite{dadonaite2024spike} \\
      SCV2 Omicron XBB.1.5& binding & 7{,}341 & \cite{dadonaite2024spike} \\
      SCV2 Omicron XBB.1.5& cell entry & 8{,}334 & \cite{dadonaite2024spike} \\
      SCV2 Omicron BA.2& binding & 7{,}498 & \cite{dadonaite2024spike} \\
      SCV2 Omicron BA.2& cell entry & 8{,}370 & \cite{dadonaite2024spike} \\
      SCV2 KP.3.11 & immune escape & 7{,}316 & \cite{dadonaite2025spike} \\
      SCV2 KP.3.11 & cell entry & 9{,}205 & \cite{dadonaite2025spike} \\
      SCV2 KP.3.11 & binding & 7{,}559 & \cite{dadonaite2025spike} \\
      *$^{\dagger}$SCV2 Wuhan hu & viral growth & 5{,}725 & \cite{flynn2022comprehensive}\\
      *$^{\dagger}$IAV PB1 & viral growth & 12{,}003 & \cite{yuan2023deep} \\
      IAV H3N2 HK19 & immune escape & 2{,}508 & \cite{welsh2024age} \\
      IAV H3N2 HK19 & viral growth & 2{,}508 & \cite{welsh2024age} \\
      $^{\dagger}$IAV H5N1& immune escape & 5{,}597 & \cite{dadonaite2024deep} \\
      $^{\dagger}$IAV H5N1& cell entry & 9{,}510 & \cite{dadonaite2024deep} \\
      $^{\dagger}$IAV H5N1& stability & 4{,}413 & \cite{dadonaite2024deep} \\
      IAV H3N2 MC22 & immune escape & 6{,}139 & \cite{yu2025pleiotropic} \\
      IAV H3N2 MC22 & cell entry & 9{,}738 & \cite{yu2025pleiotropic} \\
      IAV H3N2 MC22 & stability & 5{,}976 & \cite{yu2025pleiotropic} \\
      *$^{\dagger}$IAV H1N1 & viral growth & 10{,}715 & \cite{doud2016accurate} \\
      *$^{\dagger}$IAV H1N1 & viral growth & 2{,}350 & \cite{wu2014high} \\
      *$^{\dagger}$IAV H2N1 & viral growth & 14{,}421 & \cite{soh2019comprehensive} \\
      *$^{\dagger}$IAV H3N2 & viral growth & 10{,}754 & \cite{lee2018deep} \\
      *$^{\dagger}$IAV H3N2 & viral growth & 9{,}462 & \cite{doud2015site} \\
      *$^{\dagger}$IAV H1N1 & viral growth & 9{,}462 & \cite{doud2015site} \\
      *$^{\dagger}$IAV H1N1 & viral growth & 298 & \cite{jiang2016balance} \\
      *$^{\dagger}$IAV H1N1 & viral growth & 1{,}820 & \cite{wu2015functional} \\
      *$^{\dagger}$CVB3 & viral growth & 15{,}711 & \cite{mattenberger2021globally} \\
      *$^{\dagger}$AAV2 & viral growth & 42{,}328 & \cite{sinai2021generative} \\
      *$^{\dagger}$DEN & viral growth & 16{,}897 & \cite{suphatrakul2023functional} \\
      *HCV JFH 1 & viral growth & 1{,}630 & \cite{qi2014quantitative} \\
      *$^{\dagger}$PESV & stability & 5{,}130 & \cite{tsuboyama2023mega} \\
      \bottomrule
      \multicolumn{4}{l}{* from ProteinGym; $\dagger$ from EVEREST}
    \end{tabular}
\end{table}

\begin{table}[H]
  \caption{Per-virus breakdown of DMS entries by mutation type across all 79 assays (552,065 total entries). \emph{Substitution}: standard single amino acid change (\texttt{X\#Y}); \emph{Indel}: deletion or insertion; \emph{Combinatorial}: two or more simultaneous substitutions (colon-separated); \emph{Stop/RT}: nonsense (AA$\to$stop) or readthrough (stop$\to$AA) variants.}
  \label{tab:mut_types}
  \centering
    \begin{tabular}{lrrrrrr}
      \toprule
      Virus & Assays & Substitution & Indel & Combinatorial & Stop/RT & Total \\
      \midrule
      SARS-CoV-2  & 36 & 172{,}394 & 2{,}835 &       0 &     0 & 175{,}229 \\
      Influenza~A & 17 & 117{,}674 &       0 &       0 &     0 & 117{,}674 \\
      HIV         & 11 &  83{,}646 &     353 &       0 &    97 &  84{,}096 \\
      Nipah       &  3 &  27{,}434 &       0 &       0 &     0 &  27{,}434 \\
      Zika        &  2 &  19{,}656 &       0 &       0 &     0 &  19{,}656 \\
      Rabies      &  2 &  15{,}834 &       0 &       0 &     0 &  15{,}834 \\
      Lassa       &  2 &  12{,}754 &       0 &       0 &     0 &  12{,}754 \\
      CVB3        &  1 &  15{,}711 &       0 &       0 &     0 &  15{,}711 \\
      Dengue      &  1 &  16{,}897 &       0 &       0 &     0 &  16{,}897 \\
      HBV         &  1 &  16{,}827 &       0 &       0 &   865 &  17{,}692 \\
      HCV         &  1 &   1{,}630 &       0 &       0 &     0 &   1{,}630 \\
      AAV2        &  1 &       532 &       0 &  41{,}796 &     0 &  42{,}328 \\
      PESV        &  1 &       995 &       0 &   4{,}135 &     0 &   5{,}130 \\
      \midrule
      Total       & 79 & 501{,}984 & 3{,}188 &  45{,}931 &   962 & 552{,}065 \\
      \bottomrule
    \end{tabular}
\end{table}

\begin{table}[H]
  \caption{Distribution of viruses and phenotypes across all 79 DMS functional assays.}
  \label{detailed_types}
  \centering
    \begin{tabular}{lcccccccc}
      \toprule
       & Binding & Cell entry & Expression & Fitness & Immune escape & Stability & Viral growth & Total \\
      \midrule
      SARS-CoV-2  & 15 &  4 & 12 &   &  4 &   &  1 & 36 \\
      Influenza~A &    &  2 &    &   &  3 & 2 & 10 & 17 \\
      HIV         &    &  2 &    &   &  3 &   &  6 & 11 \\
      Nipah       &  1 &  1 &    &   &  1 &   &    &  3 \\
      Zika        &    &    &    &   &  1 &   &  1 &  2 \\
      Rabies      &    &  1 &    &   &  1 &   &    &  2 \\
      Lassa       &    &  1 &    &   &  1 &   &    &  2 \\
      CVB3        &    &    &    &   &    &   &  1 &  1 \\
      Dengue      &    &    &    &   &    &   &  1 &  1 \\
      HBV         &    &    &    & 1 &    &   &    &  1 \\
      HCV         &    &    &    &   &    &   &  1 &  1 \\
      AAV2        &    &    &    &   &    &   &  1 &  1 \\
      PESV        &    &    &    &   &    & 1 &    &  1 \\
      Total & 16 & 11 & 12 & 1 & 14 & 3 & 22 & 79 \\
      \bottomrule
    \end{tabular}
\end{table}

\subsection{Neutralisation Datasets}
\label{sec:appendix_neu}

\begin{table}[H]
  \caption{Sources of the 21 influenza A neutralisation assays used for antigenic diversity prediction. Sera from ferret antisera (post-infection) and human donors (post-vaccination) are included; subtypes H3N2 and H1N1 are represented.}
  \label{neutralisation_all_reference}
  \centering
    \begin{tabular}{lccc}
      \toprule
      Dataset & Sera Source & Subtype & Reference \\
      \midrule
      A/Massachusetts/18/2022                  & ferret & H3N2 & \cite{neu_kikawa2025near} \\
      A/Thailand/8/2022                        & ferret & H3N2 & \cite{neu_kikawa2025near} \\
      A/DistrictOfColumbia/27/2023             & ferret & H3N2 & \cite{neu_kikawa2025near} \\
      A/Croatia/10136RV/2023-egg               & ferret & H3N2 & \cite{neu_kikawa2025near} \\
      A/Netherlands/10563/2023                 & ferret & H3N2 & \cite{neu_kikawa2025near} \\
      A/Lisboa/216/2023                        & ferret & H3N2 & \cite{neu_kikawa2025near} \\
      A/Slovenia/49/2024                       & ferret & H3N2 & \cite{neu_kikawa2025near} \\
      A/Switzerland/47775/2024                 & ferret & H3N2 & \cite{neu_kikawa2025near} \\
      A/Norway/12374/2023                      & ferret & H3N2 & \cite{neu_kikawa2025near} \\
      A/BurkinaFaso/3131/2023                  & ferret & H3N2 & \cite{neu_kikawa2025near} \\
      A/France/IDF-IPP29542/2023-egg           & ferret & H3N2 & \cite{neu_kikawa2025near} \\
      A/Netherlands/10685/2024                 & ferret & H3N2 & \cite{neu_kikawa2025near} \\
      A/Lisboa/188/2023                        & ferret & H1N1 & \cite{neu_kikawa2025near} \\
      A/Victoria/4897/2022                     & ferret & H1N1 & \cite{neu_kikawa2025near} \\
      A/Victoria/4897/2022\_IVR-238            & ferret & H1N1 & \cite{neu_kikawa2025near} \\
      A/Wisconsin/67/2022                      & ferret & H1N1 & \cite{neu_kikawa2025near} \\
      A/Norway/07606/2024                      & ferret & H1N1 & \cite{neu_kikawa2025near} \\
      A/Tajikistan/02-1057/2024                & ferret & H1N1 & \cite{neu_kikawa2025near} \\
      A/Darwin/9/2021                          & human  & H3N2 & \cite{kikawa_high-throughput_2025} \\
      A/Massachusetts/18/2022                  & human  & H3N2 & \cite{kikawa_high-throughput_2025} \\
      A/Wisconsin/588/2019                     & human  & H1N1 & \cite{neu_loes2024high} \\
      \bottomrule
    \end{tabular}
\end{table}

\subsection{Indel Analysis}

Table~\ref{non_indel_types} classifies the 79 DMS tasks by mutation content. A task is \emph{substitution-only} if every entry is a standard single amino-acid change (\texttt{X\#Y}). Of the 79 tasks, 54 are strictly substitution-only, 2 contain only multi-site combinatorial mutants with no deletions (AAV2, PESV), and 23 contain deletion, insertion, stop-codon, or readthrough variants. Per-virus sequence counts by mutation type are given in Table~\ref{tab:mut_types}. Table~\ref{no_del_tasks} in Section~\ref{sec:appendix_results} reports model performance on the 56 non-deletion tasks (sub-only + combinatorial).

\begin{table}[H]
  \caption{Classification of the 79 DMS tasks by mutation content. \emph{Sub-only}: every variant is a single amino-acid substitution (\texttt{X\#Y}). \emph{Comb.}: task contains multi-site combinatorial mutants but no deletions (AAV2: 98.7\% combinatorial; PESV: 80.6\%). \emph{Indel/other}: task contains deletion, insertion, stop-codon, or readthrough variants. Per-virus sequence counts broken down by mutation type are in Table~\ref{tab:mut_types}.}
  \label{non_indel_types}
  \centering
    \begin{tabular}{lcccc}
      \toprule
      Virus & Sub-only tasks & Comb.\ tasks & Indel/other tasks & Total tasks \\
      \midrule
      SARS-CoV-2  & 19 &  0 & 17 & 36 \\
      Influenza~A & 17 &  0 &  0 & 17 \\
      HIV         &  6 &  0 &  5 & 11 \\
      Nipah       &  3 &  0 &  0 &  3 \\
      Zika        &  2 &  0 &  0 &  2 \\
      Rabies      &  2 &  0 &  0 &  2 \\
      Lassa       &  2 &  0 &  0 &  2 \\
      CVB3        &  1 &  0 &  0 &  1 \\
      Dengue      &  1 &  0 &  0 &  1 \\
      HCV         &  1 &  0 &  0 &  1 \\
      AAV2        &  0 &  1 &  0 &  1 \\
      PESV        &  0 &  1 &  0 &  1 \\
      HBV         &  0 &  0 &  1$^{*}$ &  1 \\
      \midrule
      Total  & 54 &  2 & 23 & 79 \\
      \bottomrule
      \multicolumn{5}{l}{$^{*}$HBV contains stop-codon and readthrough variants (865 entries), not actual indels.}
    \end{tabular}
\end{table}

\section{Extended Results}
\label{sec:appendix_results}

This section presents full model rankings across all 25 baselines and all tasks, supplementary analyses, and visualisations referenced in the main text.

\subsection{Full Model Rankings}
\label{sec:appendix_rankings}

Table~\ref{results_all} reports performance for all 25 baselines across DMS, neutralisation, and GISAID tasks. Table~\ref{gisaid_all_models} extends pandemic prediction to all 25,460 mutations, excluding VESPA/VESPAl because they cannot score deletions. Table~\ref{no_del_tasks} reports performance on the 56 substitution-only DMS tasks to assess the impact of excluding indels. Table~\ref{tab:dms_significance} reports pairwise significance against ProGen2-XL using exact two-sided paired sign tests with Benjamini--Hochberg correction.

\begin{table}[H]
  \caption{Results for all models across tasks.}
  \label{results_all}
  \centering
    \begin{tabular}{lcccccccr}
      \toprule
      & \multicolumn{4}{c}{DMS} & \multicolumn{2}{c}{Neutralisation} & \multicolumn{2}{c}{GISAID} \\
      Model Name & Recall & Std. & Spearman & Std. & Spearman & Std. & Recall & Spearman \\
      \midrule
      VESPAl          & 0.1681 & 0.1279 & 0.2561 & 0.1486 & 0.1965 & 0.2057 & 0.2953 & 0.3158 \\
      VESPA           & 0.1669 & 0.0971 & 0.2380 & 0.1351 & 0.1965 & 0.2057 & 0.3238 & 0.3616 \\
      Tranception S   & 0.1603 & 0.0759 & 0.2358 & 0.1381 & 0.1843 & 0.1810 & 0.2370 & 0.2894 \\
      Tranception M   & 0.1734 & 0.0755 & 0.2654 & 0.1346 & \textbf{0.2316} & 0.1696 & 0.3143 & 0.3249 \\
      Tranception L   & 0.1898 & 0.0857 & 0.2735 & 0.1482 & 0.1927 & 0.1961 & 0.3743 & 0.4014 \\
      ProtGPT2        & 0.1105 & 0.0382 & 0.1019 & 0.0768 & 0.2018 & 0.1845 & 0.0968 & 0.0495 \\
      ProGen2         & 0.1876 & 0.0848 & 0.2816 & 0.1657 & 0.2018 & 0.1929 & 0.4082 & \textbf{0.4343} \\
      ProGen2 S       & 0.1599 & 0.0675 & 0.2418 & 0.1602 & 0.2250 & 0.1852 & 0.2076 & 0.2504 \\
      ProGen2 M       & 0.1837 & 0.0825 & 0.2879 & 0.1712 & 0.2240 & 0.2070 & 0.2779 & 0.3107 \\
      ProGen2 OAS     & 0.0974 & 0.0232 & 0.0387 & 0.0367 & 0.1829 & 0.1513 & 0.1013 & 0.0822 \\
      ProGen2 BFD90   & 0.1897 & 0.0828 & 0.2789 & 0.1580 & 0.2147 & 0.1981 & 0.4078 & 0.3757 \\
      ProGen2 L       & 0.1769 & 0.0739 & 0.2618 & 0.1563 & 0.2101 & 0.2002 & 0.2738 & 0.3203 \\
      ProGen2 XL      & \textbf{0.1973} & 0.0913 & \textbf{0.2928} & 0.1595 & 0.2093 & 0.1977 & \textbf{0.4347} & 0.4018 \\
      ESM1v           & 0.1461 & 0.0667 & 0.1950 & 0.1018 & 0.2282 & 0.2043 & 0.2676 & 0.2682 \\
      ESM1 43M        & 0.1462 & 0.0659 & 0.1874 & 0.0984 & 0.2222 & 0.2098 & 0.1935 & 0.2101 \\
      ESM1 85M        & 0.1416 & 0.0525 & 0.1781 & 0.0925 & 0.2024 & 0.2217 & 0.2630 & 0.2500 \\
      ESM1 670M UR50S & 0.1466 & 0.0590 & 0.1978 & 0.1030 & 0.1957 & 0.1943 & 0.2097 & 0.1921 \\
      ESM1 670M UR50D & 0.1394 & 0.0573 & 0.1771 & 0.0942 & 0.2093 & 0.2039 & 0.2031 & 0.1933 \\
      ESM1 670M UR100 & 0.1264 & 0.0521 & 0.1054 & 0.0930 & 0.2029 & 0.1725 & 0.2370 & 0.2330 \\
      ESM2 8M         & 0.1294 & 0.0501 & 0.1388 & 0.0843 & 0.2053 & 0.1939 & 0.2560 & 0.2435 \\
      ESM2 35M        & 0.1287 & 0.0506 & 0.1401 & 0.0956 & 0.1961 & 0.1985 & 0.2667 & 0.2580 \\
      ESM2 150M       & 0.1301 & 0.0522 & 0.1141 & 0.0770 & 0.2174 & 0.2014 & 0.2448 & 0.1641 \\
      ESM2 650M       & 0.1382 & 0.0608 & 0.1678 & 0.1045 & 0.2267 & 0.1840 & 0.2423 & 0.1785 \\
      ESM2 3B         & 0.1429 & 0.0637 & 0.1659 & 0.1059 & 0.2148 & 0.1889 & 0.2440 & 0.1886 \\
      ESM2 15B        & 0.1389 & 0.0719 & 0.1730 & 0.1110 & 0.2039 & 0.1923 & 0.2676 & 0.2126 \\
      \bottomrule
    \end{tabular}
\end{table}

\begin{table}[H]
  \caption{Zero-shot pandemic prediction results for all models on all 25,460 mutations (VESPA/VESPAl excluded as they cannot score deletion mutations). Metrics: Top-10\% Recall, absolute Spearman's rank correlation, and Precision@3 against GISAID all-time mutation frequencies.}
  \label{gisaid_all_models}
  \centering
    \begin{small}
      \begin{tabular}{lccc}
        \toprule
        Model Name       & Recall & Spearman & Precision@3 \\
        \midrule
        ProGen2 XL       & \textbf{0.4077} & 0.3170 & \textbf{0.33} \\
        ProGen2 BFD90    & 0.3798 & 0.2950 & \textbf{0.33} \\
        ProGen2          & 0.3794 & 0.3245 & \textbf{0.33} \\
        Tranception L    & 0.3523 & 0.3326 & \textbf{0.33} \\
        Tranception M    & 0.2985 & 0.3217 & \textbf{0.33} \\
        ProGen2 M        & 0.2639 & 0.2428 & 0.00 \\
        ESM2 35M         & 0.2604 & 0.3033 & 0.00 \\
        ProGen2 L        & 0.2573 & 0.2535 & 0.00 \\
        ESM1 85M         & 0.2459 & 0.1877 & 0.00 \\
        ESM2 15B         & 0.2459 & 0.1862 & 0.00 \\
        ESM1v            & 0.2357 & \textbf{0.3058} & 0.00 \\
        ESM2 8M          & 0.2357 & 0.2895 & 0.00 \\
        ESM2 650M        & 0.2325 & 0.2237 & 0.00 \\
        ESM1 670M UR100  & 0.2266 & 0.2309 & 0.00 \\
        ESM2 3B          & 0.2266 & 0.2074 & 0.00 \\
        Tranception S    & 0.2207 & 0.3022 & \textbf{0.33} \\
        ESM2 150M        & 0.2121 & 0.2018 & 0.00 \\
        ESM1 670M UR50S  & 0.1952 & 0.1538 & 0.00 \\
        ProGen2 S        & 0.1948 & 0.1965 & 0.00 \\
        ESM1 670M UR50D  & 0.1925 & 0.1424 & 0.00 \\
        ESM1 43M         & 0.1866 & 0.1388 & 0.00 \\
        ProGen2 OAS      & 0.0970 & 0.0751 & 0.00 \\
        ProtGPT2         & 0.0939 & 0.0376 & 0.00 \\
        \bottomrule
      \end{tabular}
    \end{small}
\end{table}

\begin{table}[H]
  \caption{Performance on the 56 substitution-only DMS tasks (no indels). Improvement is the relative percentage change over the full 79-task results, showing modest gains from excluding indel-containing assays.}
  \label{no_del_tasks}
  \centering
    \begin{tabular}{lcccc}
      \toprule
      Model Name & Recall & Improvement (\%) & Spearman & Improvement (\%) \\
      \midrule
      ESM1 670M UR50D  & 0.1453 &  4.2070 & 0.1842 &  4.0250 \\
      ESM2 3B          & 0.1485 &  3.9177 & 0.1890 & 13.9151 \\
      VESPAl           & 0.1746 &  3.8895 & 0.2586 &  0.9669 \\
      ESM2 150M        & 0.1351 &  3.8340 & 0.1376 & 20.6038 \\
      Tranception S    & 0.1662 &  3.6926 & 0.2502 &  6.1056 \\
      Tranception L    & 0.1964 &  3.4928 & 0.2801 &  2.4001 \\
      ESM1v            & 0.1511 &  3.4565 & 0.2078 &  6.5489 \\
      ProGen2 OAS      & 0.1005 &  3.2105 & 0.0448 & 15.7688 \\
      VESPA            & 0.1715 &  2.7310 & 0.2367 & -0.5293 \\
      Tranception M    & 0.1774 &  2.2951 & 0.2704 &  1.8913 \\
      ProGen2 XL       & 0.2014 &  2.0766 & 0.2946 &  0.6008 \\
      ESM2 650M        & 0.1406 &  1.7183 & 0.1819 &  8.4272 \\
      ESM1 43M         & 0.1486 &  1.6402 & 0.1874 & -0.0254 \\
      ProGen2 L        & 0.1796 &  1.5195 & 0.2659 &  1.5615 \\
      ESM2 8M          & 0.1313 &  1.4925 & 0.1510 &  8.7762 \\
      ESM2 15B         & 0.1409 &  1.4734 & 0.1767 &  2.1137 \\
      ProGen2 M        & 0.1861 &  1.2998 & 0.2868 & -0.3670 \\
      ESM2 35M         & 0.1300 &  0.9989 & 0.1543 & 10.1291 \\
      ESM1 85M         & 0.1427 &  0.7791 & 0.1806 &  1.3813 \\
      ProGen2 BFD90    & 0.1905 &  0.4439 & 0.2761 & -0.9917 \\
      ESM1 670M UR50S  & 0.1465 & -0.0707 & 0.2020 &  2.1315 \\
      ESM1 670M UR100  & 0.1263 & -0.0828 & 0.0975 & -7.5306 \\
      ProtGPT2         & 0.1086 & -1.6921 & 0.0913 & -10.3577 \\
      ProGen2 S        & 0.1570 & -1.8252 & 0.2362 & -2.3304 \\
      ProGen2          & 0.1832 & -2.3442 & 0.2733 & -2.9375 \\
      \bottomrule
    \end{tabular}
\end{table}

\begin{table}[H]
  \caption{Pairwise statistical significance of DMS benchmark comparisons against ProGen2-XL.
    For each model, $\Delta$ is the mean difference (model $-$ ProGen2-XL) across all 79 shared
    tasks. $p_{\mathrm{adj}}$ is the two-sided exact paired sign-test $p$-value after
    Benjamini--Hochberg correction within each metric ($n=79$ tasks for all pairs).
    Significance: $^{***}p<0.001$, $^{**}p<0.01$, $^{*}p<0.05$, $^{\mathrm{ns}}$not significant ($\alpha=0.05$).}
  \label{tab:dms_significance}
  \centering
  \begin{small}
    \begin{tabular}{lrrrrrr}
      \toprule
      & \multicolumn{3}{c}{Top-10\% Recall} & \multicolumn{3}{c}{$|\mathrm{Spearman}|$} \\
      \cmidrule(lr){2-4}\cmidrule(lr){5-7}
      Model & Mean & $\Delta$ & $p_{\mathrm{adj}}$ & Mean & $\Delta$ & $p_{\mathrm{adj}}$ \\
      \midrule
      \textbf{ProGen2 XL} (ref) & \textbf{0.1973} & --- & --- & \textbf{0.2928} & --- & --- \\
      \midrule
      VESPAl                        & 0.1682 & $-$0.0291 & $0.0001^{***}$ & 0.2561 & $-$0.0367 & $0.0510^{\mathrm{ns}}$ \\
      VESPA                         & 0.1669 & $-$0.0304 & $0.0002^{***}$ & 0.2380 & $-$0.0548 & $0.0000^{***}$ \\
      Tranception S                 & 0.1603 & $-$0.0370 & $0.0004^{***}$ & 0.2357 & $-$0.0571 & $0.0043^{**}$ \\
      Tranception M                 & 0.1734 & $-$0.0239 & $0.0002^{***}$ & 0.2653 & $-$0.0275 & $0.0043^{**}$ \\
      Tranception L                 & 0.1898 & $-$0.0075 & $0.5466^{\mathrm{ns}}$ & 0.2735 & $-$0.0193 & $0.0021^{**}$ \\
      ProtGPT2                      & 0.1105 & $-$0.0868 & $0.0000^{***}$ & 0.1019 & $-$0.1909 & $0.0000^{***}$ \\
      ProGen2 Base                  & 0.1876 & $-$0.0097 & $0.6249^{\mathrm{ns}}$ & 0.2816 & $-$0.0112 & $1.0000^{\mathrm{ns}}$ \\
      ProGen2 S                     & 0.1599 & $-$0.0374 & $0.0138^{*}$   & 0.2417 & $-$0.0511 & $0.0043^{**}$ \\
      ProGen2 M                     & 0.1837 & $-$0.0136 & $0.0191^{*}$   & 0.2879 & $-$0.0049 & $0.0510^{\mathrm{ns}}$ \\
      ProGen2 OAS                   & 0.0974 & $-$0.0999 & $0.0000^{***}$ & 0.0387 & $-$0.2541 & $0.0000^{***}$ \\
      ProGen2 BFD90                 & 0.1897 & $-$0.0076 & $0.1055^{\mathrm{ns}}$ & 0.2789 & $-$0.0139 & $0.1293^{\mathrm{ns}}$ \\
      ProGen2 L                     & 0.1770 & $-$0.0203 & $0.0004^{***}$ & 0.2618 & $-$0.0310 & $0.1293^{\mathrm{ns}}$ \\
      ESM1v                         & 0.1461 & $-$0.0512 & $0.0000^{***}$ & 0.1950 & $-$0.0978 & $0.0000^{***}$ \\
      ESM1 43M                      & 0.1462 & $-$0.0511 & $0.0000^{***}$ & 0.1873 & $-$0.1055 & $0.0000^{***}$ \\
      ESM1 85M                      & 0.1416 & $-$0.0557 & $0.0000^{***}$ & 0.1781 & $-$0.1147 & $0.0000^{***}$ \\
      ESM1 670M UR50S               & 0.1467 & $-$0.0506 & $0.0000^{***}$ & 0.1978 & $-$0.0950 & $0.0000^{***}$ \\
      ESM1 670M UR50D               & 0.1394 & $-$0.0579 & $0.0000^{***}$ & 0.1771 & $-$0.1157 & $0.0000^{***}$ \\
      ESM1 670M UR100               & 0.1265 & $-$0.0708 & $0.0000^{***}$ & 0.1054 & $-$0.1874 & $0.0000^{***}$ \\
      ESM2 8M                       & 0.1295 & $-$0.0678 & $0.0000^{***}$ & 0.1387 & $-$0.1541 & $0.0000^{***}$ \\
      ESM2 35M                      & 0.1287 & $-$0.0686 & $0.0000^{***}$ & 0.1401 & $-$0.1527 & $0.0000^{***}$ \\
      ESM2 150M                     & 0.1301 & $-$0.0672 & $0.0000^{***}$ & 0.1141 & $-$0.1787 & $0.0000^{***}$ \\
      ESM2 650M                     & 0.1382 & $-$0.0591 & $0.0000^{***}$ & 0.1678 & $-$0.1250 & $0.0000^{***}$ \\
      ESM2 3B                       & 0.1429 & $-$0.0544 & $0.0000^{***}$ & 0.1659 & $-$0.1269 & $0.0000^{***}$ \\
      ESM2 15B                      & 0.1389 & $-$0.0584 & $0.0000^{***}$ & 0.1729 & $-$0.1199 & $0.0000^{***}$ \\
      \bottomrule
    \end{tabular}
  \end{small}
\end{table}

\subsection{Cross-model DMS--Pandemic Correlation}

Table~\ref{tab:dms_pandemic_corr} reports the Spearman correlation across all 25 baselines between DMS performance and SARS-CoV-2 all-time GISAID pandemic prediction. All four comparisons are highly significant, providing direct evidence that DMS performance predicts real-world pandemic performance. DMS can therefore serve as a low-cost screen for selecting pLMs for mutation prioritisation. The top-3 models by DMS Recall (ProGen2-XL, Tranception-L, ProGen2-BFD90) are also the top-3 by pandemic Recall.

\begin{table}[H]
  \caption{Cross-model Spearman correlation ($n=25$ baselines) between DMS and pandemic-prediction rankings.}
  \label{tab:dms_pandemic_corr}
  \centering
  \begin{small}
    \begin{tabular}{lcc}
      \toprule
      Comparison & $\rho$ ($n=25$) & $p$ \\
      \midrule
      DMS Recall vs.\ Pandemic Recall     & 0.753 & $1.4\times10^{-5}$ \\
      DMS Recall vs.\ Pandemic $|\rho|$    & 0.858 & $4.1\times10^{-8}$ \\
      DMS $|\rho|$ vs.\ Pandemic Recall    & 0.736 & $2.7\times10^{-5}$ \\
      DMS $|\rho|$ vs.\ Pandemic $|\rho|$  & 0.862 & $3.2\times10^{-8}$ \\
      \bottomrule
    \end{tabular}
  \end{small}
\end{table}

\subsection{Signed Spearman Correlation}
\label{sec:appendix_signed}

The main text adopts the absolute Spearman convention of \citet{meier2021language} to accommodate the mixed directionality of different DMS assays and scoring strategies. To confirm that this choice does not affect the model ranking, we apply a per-family sign convention (Table~\ref{tab:sign_convention}) so that ``higher score $=$ higher fitness'' and recompute signed Spearman across all 25 baselines (Table~\ref{tab:signed_spearman}). Ranking all 25 baselines by mean signed $\rho$ versus mean $|\rho|$ gives a Spearman rank correlation of $0.997$ ($p<10^{-26}$) and Kendall $\tau=0.973$; the top-3, top-5, and top-9 models are identical in both orderings, and the only movement within the top-10 is a swap at ranks 10--11 between Tranception~S and VESPA, which differ by less than $0.005$ in both metrics. The ``\% correct direction'' column additionally identifies ProGen2-OAS as the sole directionally-random outlier ($51.9\%$), consistent with its near-zero correlations throughout.

\begin{table}[H]
  \caption{Per-family sign convention applied before computing signed Spearman correlation.}
  \label{tab:sign_convention}
  \centering
  \begin{small}
    \begin{tabular}{llll}
      \toprule
      Model family & Score column & Direction & Sign \\
      \midrule
      ESM1, ESM1v, ESM2 (semantic) & \texttt{euc\_dis} & higher $=$ further from WT $=$ less fit & negate \\
      VESPA / VESPAl               & \texttt{VESPA}/\texttt{VESPAl} & higher $=$ more pathogenic $=$ less fit & negate \\
      ProGen2 (all sizes)          & \texttt{ll} & higher $=$ more likely $=$ more fit & keep \\
      Tranception S/M/L            & \texttt{avg\_score} & higher $=$ more fit & keep \\
      ProtGPT2                     & \texttt{ProtGPT2\_score} & higher $=$ more fit & keep \\
      \bottomrule
    \end{tabular}
  \end{small}
\end{table}

\begin{table}[H]
  \caption{Signed Spearman results for all 25 baselines, ordered by mean signed $\rho$. ``\% correct direction'' is the fraction of the 79 DMS assays on which the signed correlation has the expected sign. The ordering matches the absolute-Spearman leaderboard with rank correlation $\rho=0.997$ (Kendall $\tau=0.973$); ProGen2-OAS is the sole directionally-random outlier.}
  \label{tab:signed_spearman}
  \centering
  \begin{small}
    \begin{tabular}{lccc}
      \toprule
      Model & Mean signed $\rho$ & Mean $|\rho|$ & \% correct direction \\
      \midrule
      ProGen2 XL      & 0.267 & 0.291 & 86.1 \\
      ProGen2 M       & 0.258 & 0.286 & 84.8 \\
      ProGen2         & 0.253 & 0.281 & 84.8 \\
      ProGen2 BFD90   & 0.252 & 0.277 & 82.3 \\
      Tranception L   & 0.244 & 0.274 & 86.1 \\
      Tranception M   & 0.236 & 0.265 & 83.5 \\
      ProGen2 L       & 0.233 & 0.260 & 86.1 \\
      VESPAl          & 0.222 & 0.256 & 82.3 \\
      ProGen2 S       & 0.214 & 0.239 & 86.1 \\
      Tranception S   & 0.212 & 0.236 & 83.5 \\
      VESPA           & 0.207 & 0.238 & 86.1 \\
      ESM1v           & 0.180 & 0.195 & 93.7 \\
      ESM1 670M UR50S & 0.179 & 0.198 & 86.1 \\
      ESM1 43M        & 0.171 & 0.187 & 91.1 \\
      ESM1 85M        & 0.164 & 0.178 & 91.1 \\
      ESM1 670M UR50D & 0.162 & 0.177 & 88.6 \\
      ESM2 650M       & 0.151 & 0.168 & 91.1 \\
      ESM2 15B        & 0.149 & 0.173 & 81.0 \\
      ESM2 3B         & 0.146 & 0.166 & 88.6 \\
      ESM2 35M        & 0.125 & 0.140 & 91.1 \\
      ESM2 8M         & 0.122 & 0.139 & 89.9 \\
      ESM2 150M       & 0.101 & 0.114 & 87.3 \\
      ProtGPT2        & 0.086 & 0.101 & 79.7 \\
      ESM1 670M UR100 & 0.083 & 0.105 & 74.7 \\
      ProGen2 OAS     & 0.005 & 0.039 & 51.9 \\
      \bottomrule
    \end{tabular}
  \end{small}
\end{table}

\subsection{Training-Data Coverage and Leakage Mitigation}

Table~\ref{tab:leakage} documents, per baseline family, the training corpus, documented training cutoff or release date, whether viral sequences were retained during pre-training, whether post-2021 SARS-CoV-2/GISAID-era sequences could overlap the pandemic-prediction window, and the action taken to avoid leakage. Because most ViroGym evaluation data were collected after 2020, we restrict the zero-shot pandemic-prediction leaderboard to models with documented pre-2021 training cutoffs; models whose training data include post-2021 SARS-CoV-2/GISAID sequences (e.g.\ ESM3) are excluded from the zero-shot leaderboard and reserved for a separate leakage-aware analysis.

\begin{table}[H]
  \caption{Training-data coverage and leakage-mitigation status per baseline family. ``Overlap risk'' indicates whether post-2021 SARS-CoV-2/GISAID sequences could overlap the pandemic-prediction window.}
  \label{tab:leakage}
  \centering
  \begin{small}
    \begin{tabular}{lllccl}
      \toprule
      Model & Training corpus & Cutoff/release & Viral retained & Overlap risk & Action \\
      \midrule
      ESM1        & UniRef                   & 2019       & Yes & No  & Included \\
      ESM1v       & UniRef                   & 2019       & Yes & No  & Included \\
      ESM2        & UniRef                   & Apr 2021   & Yes & No  & Included \\
      ProGen2     & BFD30 / UniRef90         & pub.\ 2023 & Yes & No  & Included \\
      Tranception & UniRef100                & Apr 2021   & Yes & No  & Included \\
      VESPA/VESPAl& ProtT5-XL / ProtBERT-BFD & $\le$2021  & Yes & No  & Included \\
      ProtGPT2    & UniRef50                 & 2021       & Yes & No  & Included \\
      \midrule
      ESM3        & incl.\ post-2021 GISAID  & 2024       & Yes & Yes & Excluded from zero-shot \\
      \bottomrule
    \end{tabular}
  \end{small}
\end{table}

\subsection{Variant-Weighted DMS Aggregation}

The main DMS results (Table~\ref{DMS}) use a macro-mean, weighting each assay equally regardless of variant count. Table~\ref{tab:dms_weighted} reports an alternative in which each assay's contribution is proportional to its number of variants. Rankings are essentially unchanged: ProGen2-XL remains first on both metrics and the top models are preserved, confirming robustness to aggregation choice.

\begin{table}[H]
  \caption{Macro-mean vs.\ variant-weighted DMS aggregation across the 79 assays. ``w'' columns weight each assay by its variant count; macro columns weight each assay equally (as in Table~\ref{DMS}).}
  \label{tab:dms_weighted}
  \centering
  \begin{small}
    \begin{tabular}{lcccc}
      \toprule
      Model & Recall (macro) & Recall (w) & $|\rho|$ (macro) & $|\rho|$ (w) \\
      \midrule
      ProGen2 XL      & 0.1957 & 0.1989 & 0.2911 & 0.3010 \\
      Tranception L   & 0.1898 & 0.1975 & 0.2735 & 0.2993 \\
      ProGen2 BFD90   & 0.1874 & 0.1828 & 0.2768 & 0.2625 \\
      ProGen2         & 0.1855 & 0.1826 & 0.2808 & 0.2819 \\
      ProGen2 M       & 0.1813 & 0.1776 & 0.2862 & 0.2700 \\
      ProGen2 L       & 0.1748 & 0.1755 & 0.2599 & 0.2577 \\
      Tranception M   & 0.1734 & 0.1744 & 0.2654 & 0.2684 \\
      VESPA           & 0.1669 & 0.1652 & 0.2380 & 0.2400 \\
      Tranception S   & 0.1603 & 0.1594 & 0.2358 & 0.2360 \\
      VESPAl          & 0.1681 & 0.1548 & 0.2561 & 0.2337 \\
      ESM1 670M UR50S & 0.1466 & 0.1510 & 0.1978 & 0.2034 \\
      ProGen2 S       & 0.1582 & 0.1506 & 0.2394 & 0.2235 \\
      ESM1 43M        & 0.1462 & 0.1482 & 0.1874 & 0.1884 \\
      ESM1 670M UR50D & 0.1394 & 0.1466 & 0.1771 & 0.1928 \\
      ESM1v           & 0.1461 & 0.1452 & 0.1950 & 0.1902 \\
      ESM1 85M        & 0.1416 & 0.1436 & 0.1781 & 0.1837 \\
      ESM2 3B         & 0.1429 & 0.1390 & 0.1659 & 0.1626 \\
      ESM2 650M       & 0.1382 & 0.1343 & 0.1678 & 0.1524 \\
      ESM2 15B        & 0.1389 & 0.1342 & 0.1730 & 0.1703 \\
      ESM2 8M         & 0.1294 & 0.1332 & 0.1388 & 0.1458 \\
      ESM2 35M        & 0.1287 & 0.1324 & 0.1401 & 0.1389 \\
      ESM2 150M       & 0.1301 & 0.1310 & 0.1141 & 0.1200 \\
      ESM1 670M UR100 & 0.1264 & 0.1295 & 0.1054 & 0.1152 \\
      ProtGPT2        & 0.1099 & 0.1054 & 0.1014 & 0.0959 \\
      ProGen2 OAS     & 0.0976 & 0.1002 & 0.0389 & 0.0449 \\
      \bottomrule
    \end{tabular}
  \end{small}
\end{table}

\subsection{Per-Task and Pandemic Figures}
\label{sec:appendix_figures}

\begin{figure*}[ht]
  \centering
    \centerline{\includegraphics[width=\textwidth]{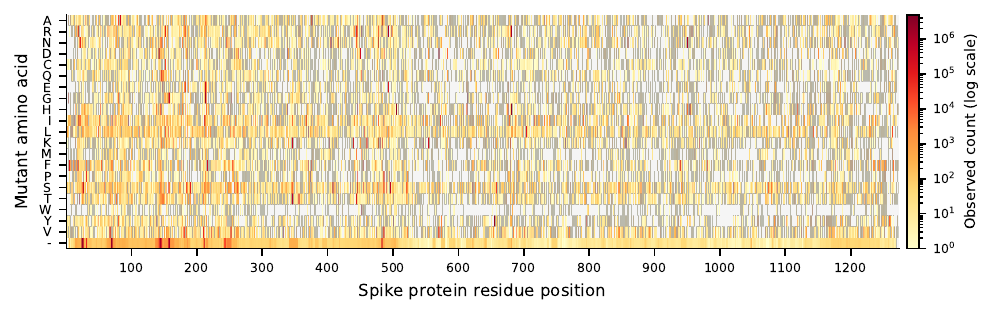}}
    \caption{
    SARS-CoV-2 Spike mutation heat map. The heat map shows the frequency of 21 potential amino acid substitutions across 1273 residues, with colour intensity indicating mutation frequency at each position. Data were collected from GISAID between January 2020 and May 2025.
    }
    \label{heatmap_GISAID}
\end{figure*}

\begin{figure*}[ht]

  \centering
    \centerline{\includegraphics[width=\textwidth]{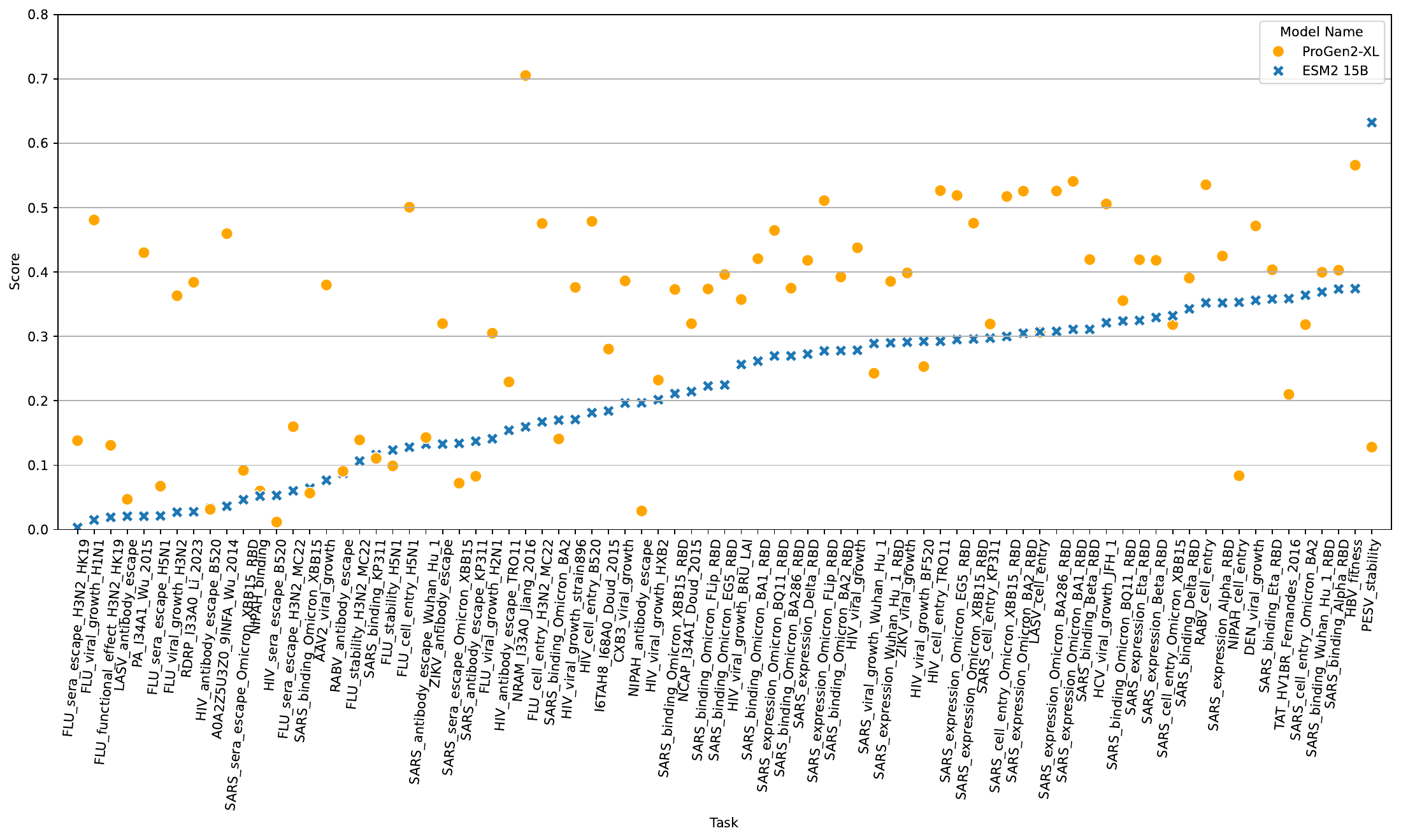}}
    \caption{
      Task-wise comparison of ESM2 15B and ProGen2-XL on the DMS benchmark. ESM2 15B uses semantic scoring, while ProGen2-XL uses negative log-likelihood. Values are absolute Spearman rank correlations between model fitness scores and experimental measurements.
    }
    \label{DMS_PER_TASK}
\end{figure*}

\begin{figure*}[ht]

  \centering
    \centerline{\includegraphics[width=\textwidth]{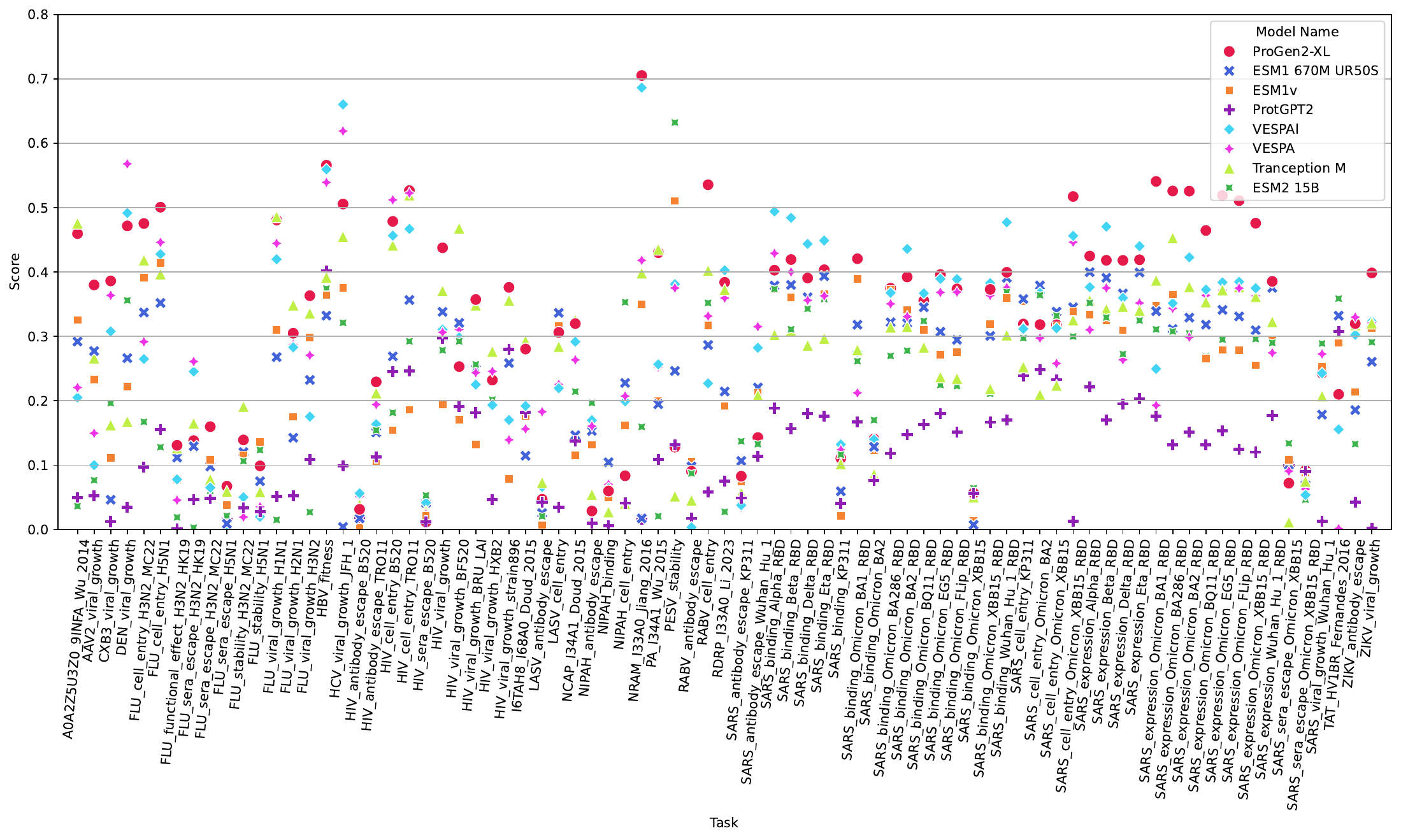}}
    \caption{
      Task-wise comparison of all baselines on the DMS benchmark. Reported values represent the absolute Spearman's rank correlation between model fitness scores and experimental measurements.
    }
    \label{dms_all_r_per_task}
\end{figure*}

\begin{figure*}[ht]

  \centering
    \centerline{\includegraphics[width=\textwidth]{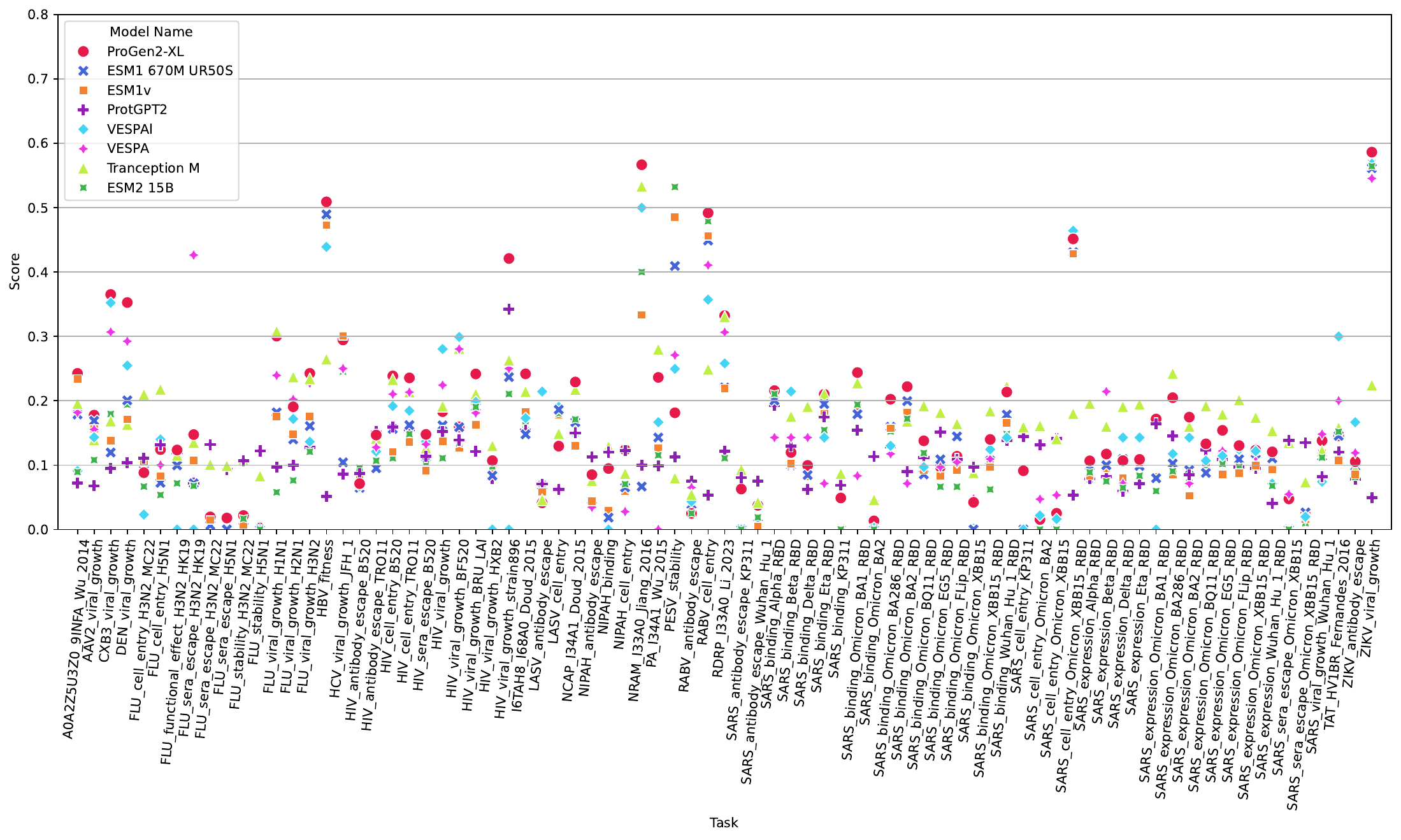}}
    \caption{
      Task-wise comparison of all baselines on the DMS benchmark. Reported values represent the top 10\% recall between model fitness scores and experimental measurements.
    }
    \label{dms_all_per_task}
\end{figure*}

\begin{figure*}[ht]

  \centering
    \centerline{\includegraphics[width=\textwidth]{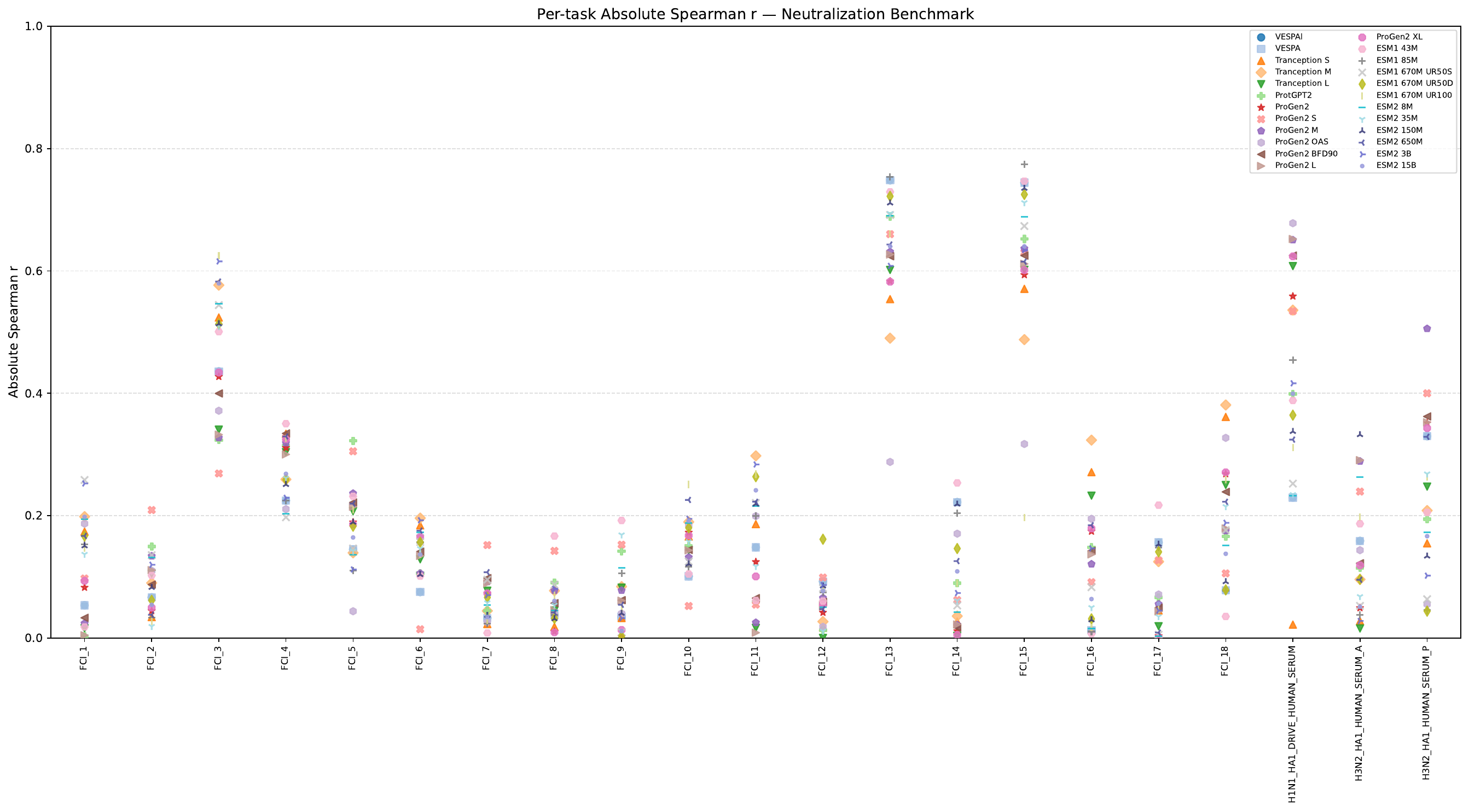}}
    \caption{
      Task-wise comparison of all baselines on the neutralisation benchmark. Reported values represent the absolute Spearman's rank correlation between model fitness scores and experimental measurements.
    }
    \label{neu_all_per_task}
\end{figure*}

\begin{figure*}[ht]

  \centering
    \centerline{\includegraphics[width=\textwidth]{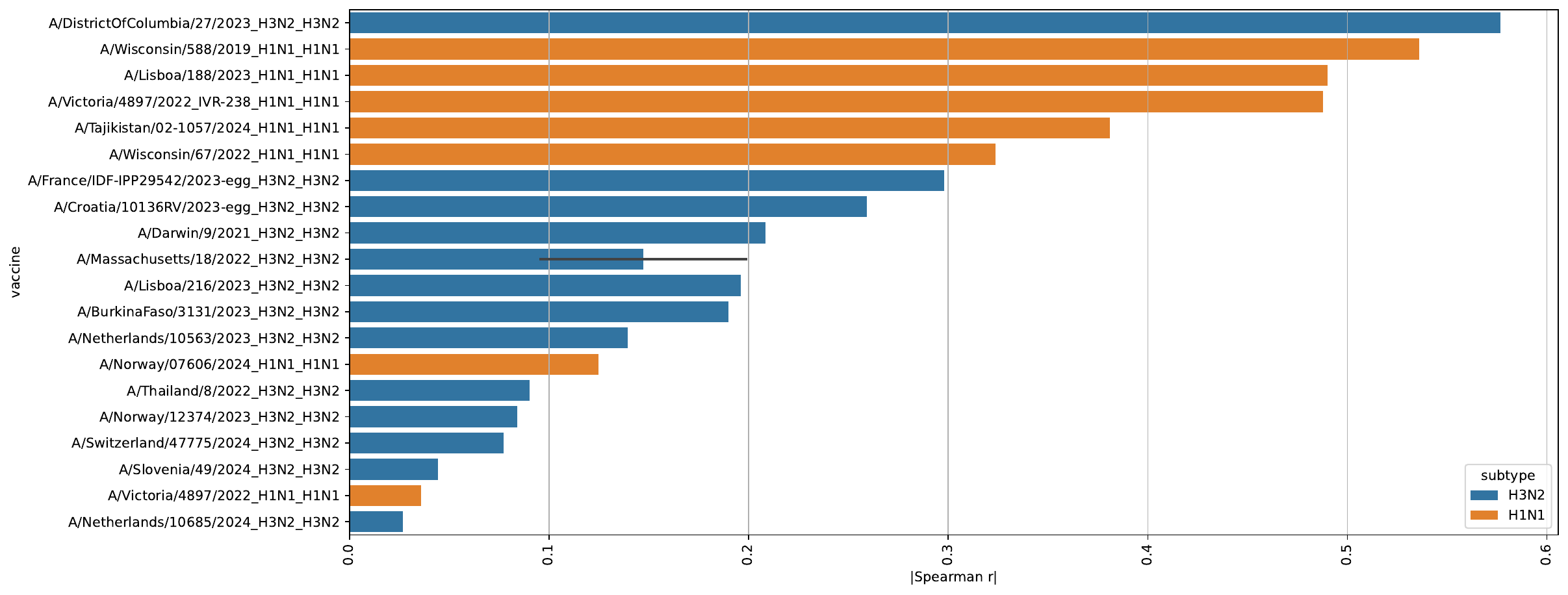}}
    \caption{
      Task-wise performance of Tranception M on the neutralisation benchmark. Antigenicity scores use negative log-likelihood. Each task corresponds to a vaccine strain representing post-vaccination serum, with colours indicating influenza A subtypes. Performance is the absolute Spearman rank correlation between predicted antigenicity scores and experimental measurements, averaged across sera from different animal sources.
    }
    \label{TM_AN}
\end{figure*}

\begin{figure}[ht]

  \centering
    \includegraphics[width=0.6\textwidth]{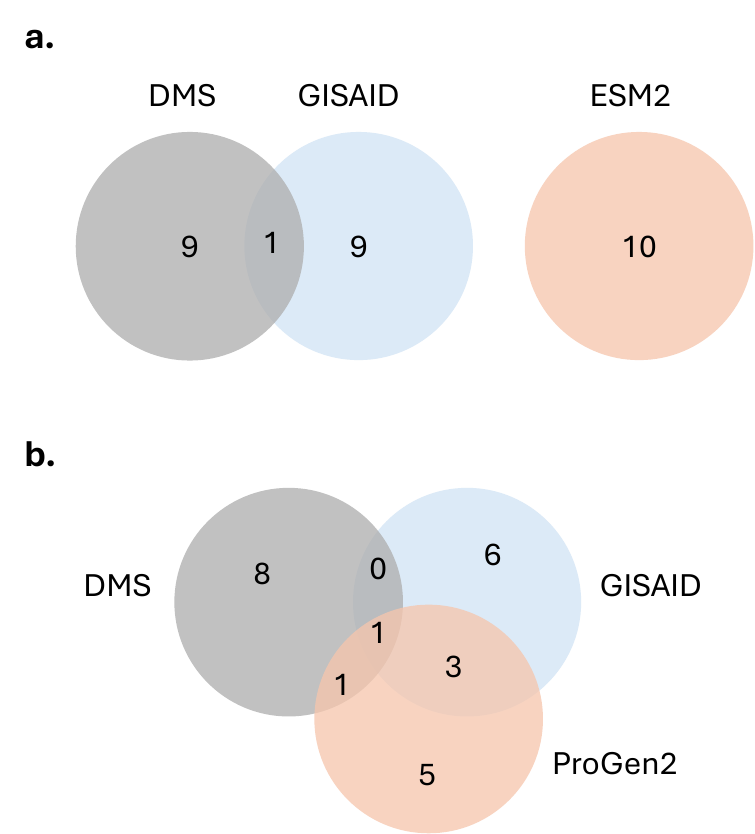}
    \caption{Overlap among top-10 RBD mutations (Spike positions 331--531) from computational predictions, in vitro DMS assays (Wuhan-Hu-1 ACE2 binding), and naturally occurring mutations from GISAID. \textbf{a.}~ESM2-650M predictions show no overlap with DMS or GISAID mutations. \textbf{b.}~ProGen2-XL overlaps $\sim$40\% with GISAID and $\sim$20\% with DMS, while DMS and GISAID share only $\sim$10\%, confirming that pLM and DMS signals are complementary.}
    \label{LINK}
\end{figure}

The following figures provide task-level performance breakdowns and GISAID visualisations referenced in the main text. Figure~\ref{heatmap_GISAID} shows observed mutation frequencies across SARS-CoV-2 Spike from GISAID (2020--2025). Figures~\ref{DMS_PER_TASK}--\ref{TM_AN} show per-task DMS and neutralisation performance, while Figure~\ref{LINK} shows top-10 RBD mutation overlap among pLM predictions, DMS assays, and GISAID.

\subsection{In Silico Score Heatmaps}
\label{sec:heatmaps}

\begin{figure*}[ht]
  \centering
    \centerline{\includegraphics[width=\textwidth]{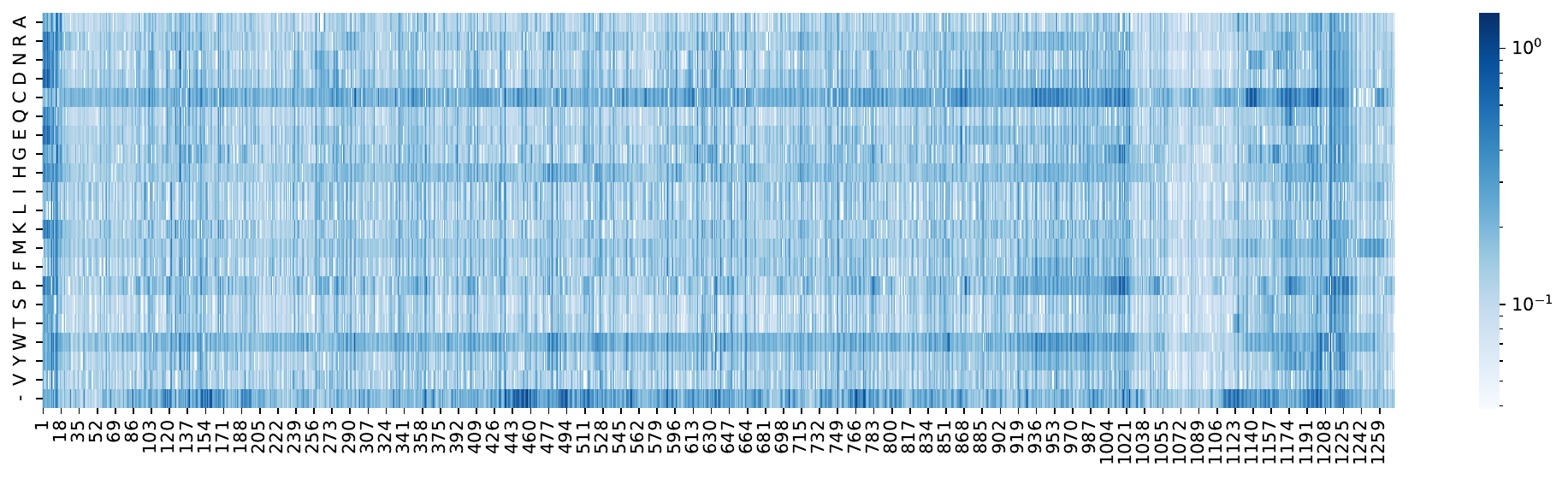}}
    \caption{In silico fitness score heatmap for ESM1. Colour intensity indicates the predicted mutational effect (semantic distance score) at each position.}
    \label{heatmap_esm1}
\end{figure*}

\begin{figure*}[ht]

  \centering
    \centerline{\includegraphics[width=\textwidth]{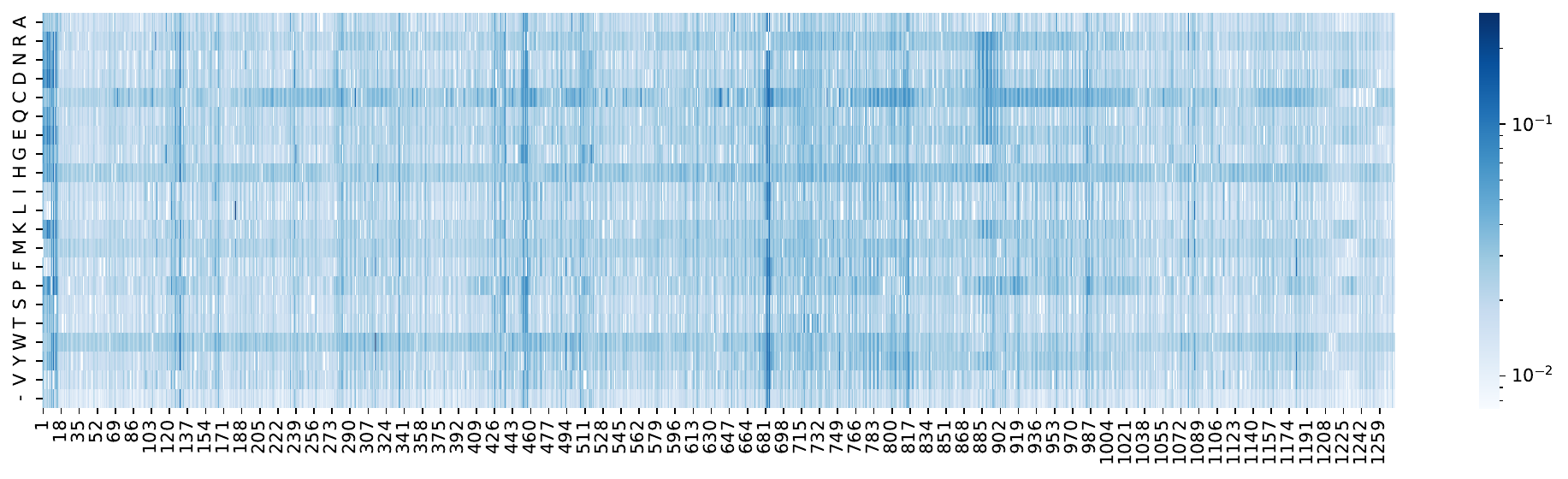}}
    \caption{
      In silico fitness score heatmap for ESM2 (semantic distance scoring).
    }
    \label{heatmap_esm2}
\end{figure*}

\begin{figure*}[ht]

  \centering
    \centerline{\includegraphics[width=\textwidth]{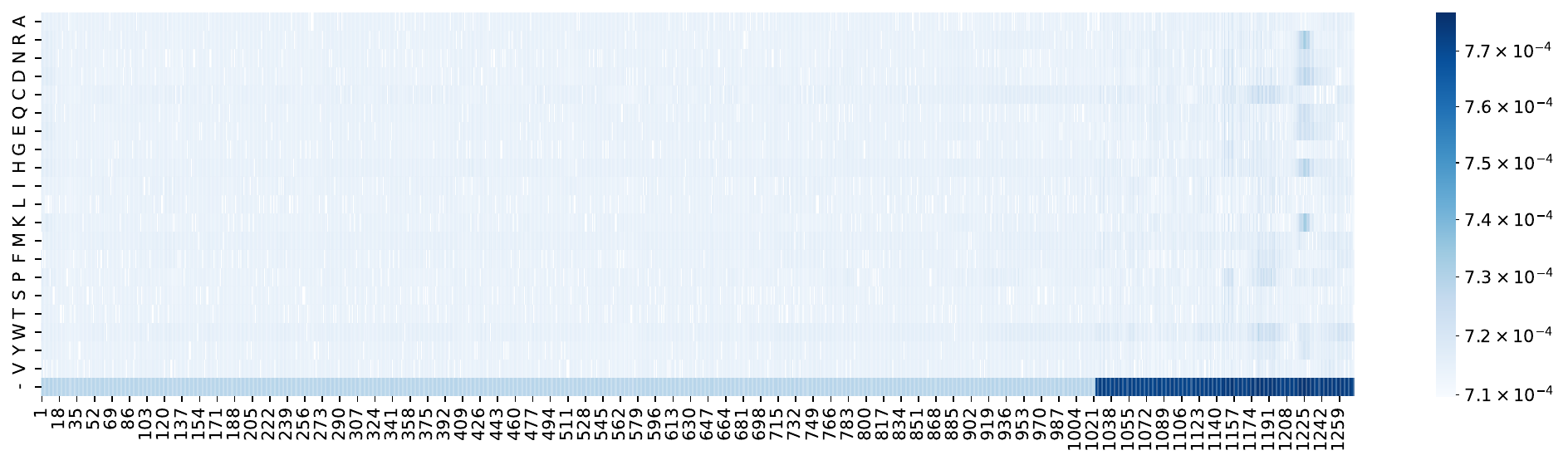}}
    \caption{
      In silico fitness score heatmap for ESM1v (semantic distance scoring).
    }
    \label{heatmap_esm1v}
\end{figure*}

\begin{figure*}[ht]

  \centering
    \centerline{\includegraphics[width=\textwidth]{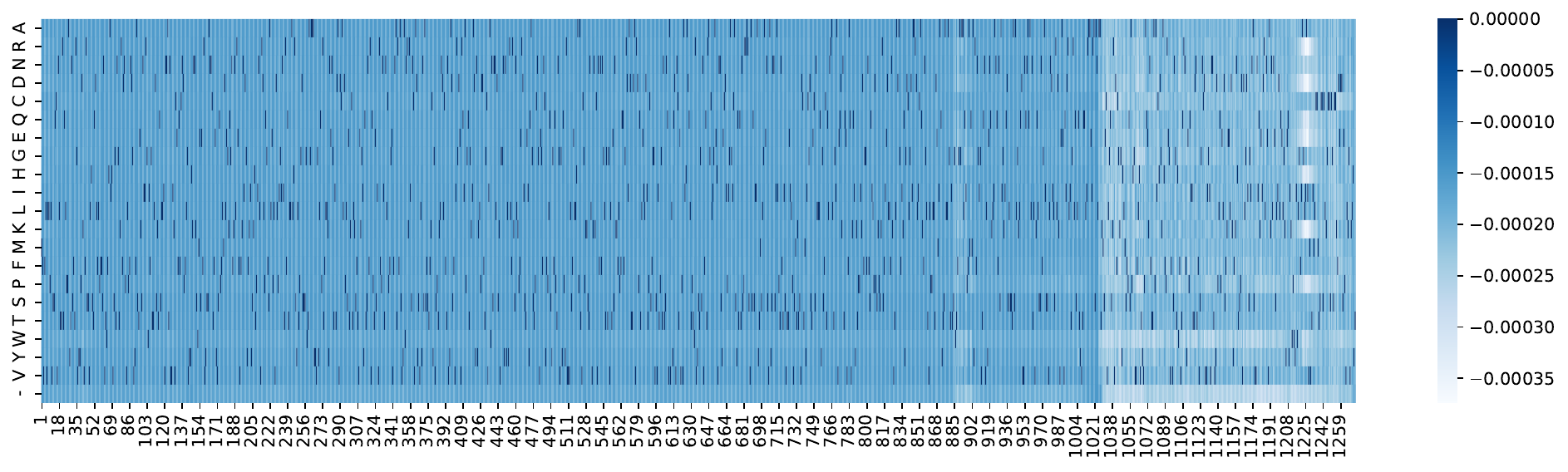}}
    \caption{
      In silico fitness score heatmap for ProGen2 (negative log-likelihood scoring).
    }
    \label{heatmap_progen2}
\end{figure*}

\begin{figure*}[ht]

  \centering
    \centerline{\includegraphics[width=\textwidth]{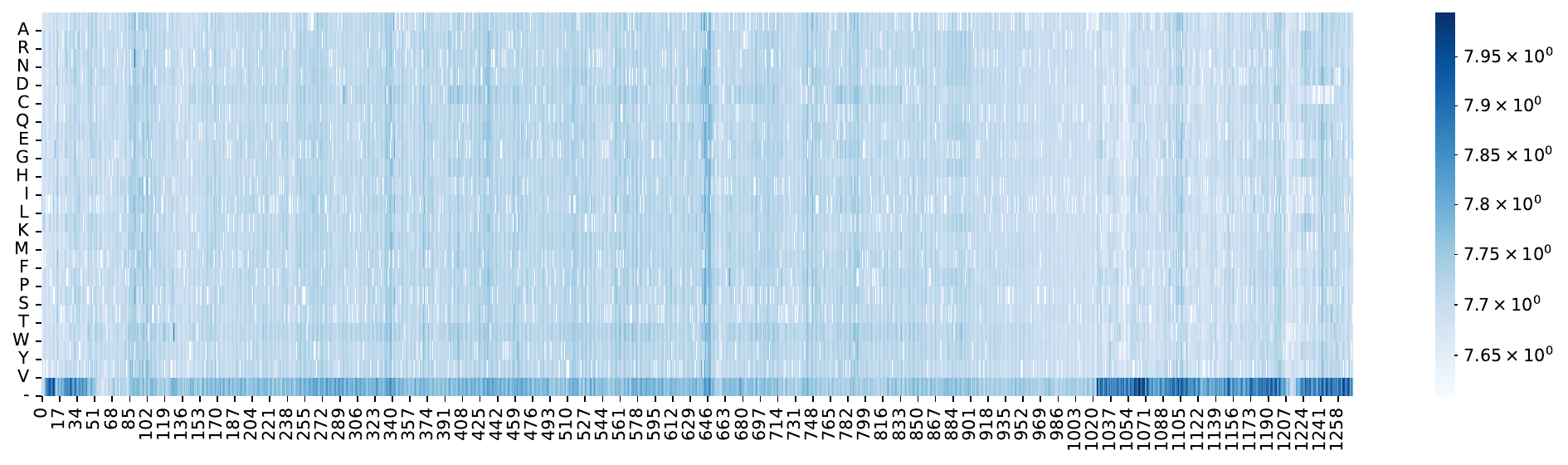}}
    \caption{
      In silico fitness score heatmap for ProtGPT2 (negative log-likelihood scoring).
    }
    \label{heatmap_ProtGPT2}
\end{figure*}

\begin{figure*}[ht]

  \centering
    \centerline{\includegraphics[width=\textwidth]{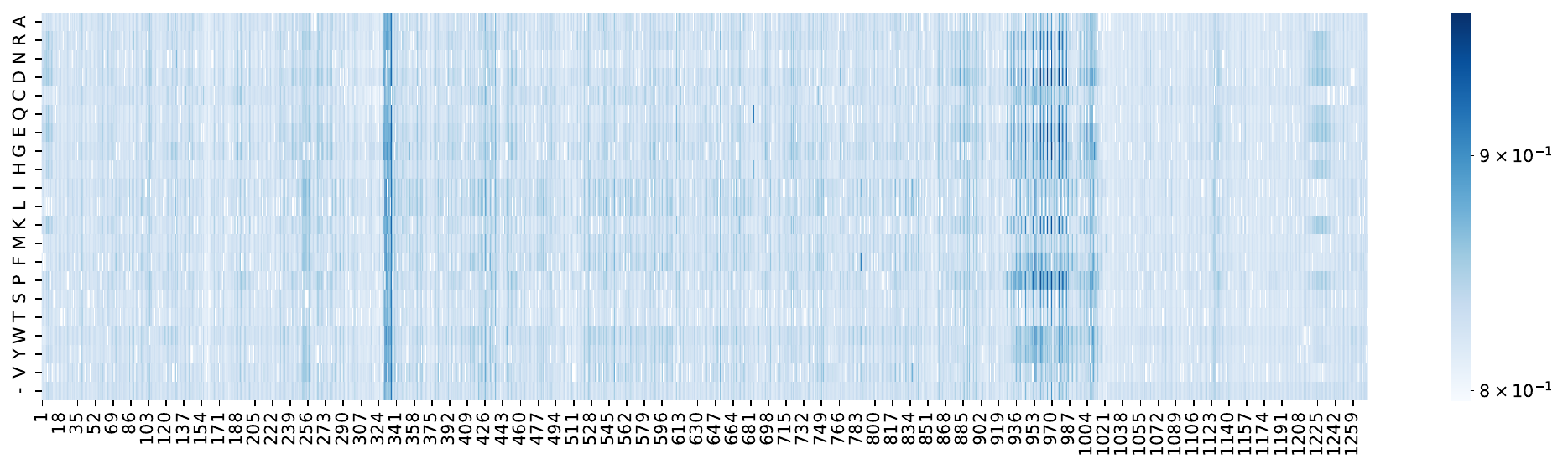}}
    \caption{
      In silico fitness score heatmap for Tranception (negative log-likelihood scoring).
    }
    \label{heatmap_tran}
\end{figure*}

\begin{figure*}[ht]

  \centering
    \centerline{\includegraphics[width=\textwidth]{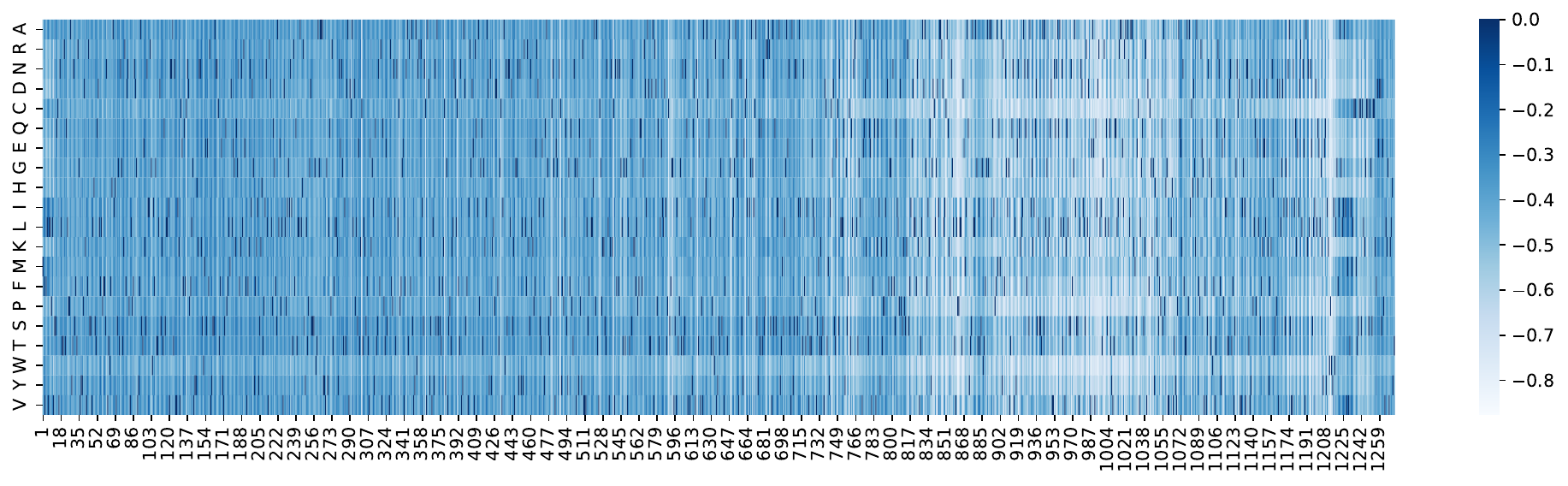}}
    \caption{
      In silico fitness score heatmap for VESPA (euclidean distance scoring).
    }
    \label{heatmap_vespa}
\end{figure*}

\begin{figure*}[ht]

  \centering
    \centerline{\includegraphics[width=\textwidth]{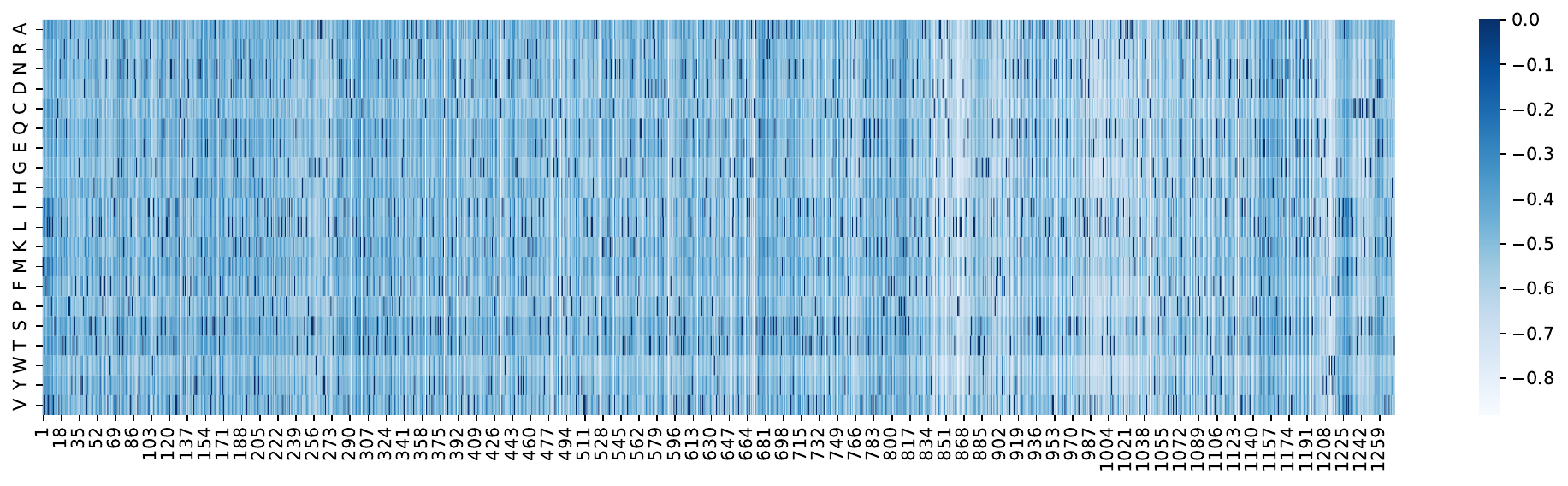}}
    \caption{
      In silico fitness score heatmap for VESPAl (euclidean distance scoring).
    }
    \label{heatmap_vespal}
\end{figure*}

Each heatmap shows the predicted in silico fitness score for every single amino acid substitution across all 1,273 residues of the SARS-CoV-2 Spike protein, computed using the best-performing scoring strategy for each model family (semantic distance for encoder-based models; negative log-likelihood for decoder-based models). Warmer colours indicate higher predicted fitness. These figures are referenced in the main text as Figures~\ref{heatmap_esm1}--\ref{heatmap_vespal}.

\clearpage
\section{Processing Pipeline and Quality Controls}
\label{sec:appendix_processing}

\paragraph{DMS.}
All 79 DMS assays were exported to a canonical three-column schema: \texttt{mutation} (single-letter substitution code, e.g.\ \texttt{N501Y}), \texttt{mutated\_sequence} (full protein with the substitution applied), and \texttt{DMS\_score} (continuous fitness proxy).
Virus-specific exporters remap each raw dataset to this schema.
For SARS-CoV-2 RBD binding and expression assays, measurements are restricted to Spike residues 331--531 (receptor-binding domain).
For HIV broadly-neutralising-antibody (MAb) escape assays, published site maps convert reference-relative positions to sequential indices, and scores measured under two overlapping antibody panels are averaged per mutation.
Datasets with multiple biological replicates are collapsed by taking the arithmetic mean over replicate scores; variants carrying a self-substitution (wild-type residue unchanged) are excluded.
The 24 datasets overlapping with ProteinGym are imported directly via the ProteinGym exporter using published substitution-level scores; multi-site combinatorial and indel variants present in raw sources are retained as-is, consistent with the original study designs (see Table~\ref{non_indel_types}).

\paragraph{Neutralisation.}
All 21 influenza~A neutralisation tasks share a three-column schema: \texttt{virus} (strain name), \texttt{titer} (HI log$_2$ titre), and \texttt{sequence} (haemagglutinin protein sequence).
For the 18 ferret-serum tasks, HA sequences are drawn from a reference FASTA of recent HA ectodomains; for 7 strains where this FASTA contained ambiguous or missing residues, sequences were manually verified and corrected.
Each benchmark row is paired with the immunising-strain sequence, which models use as their reference to score antigenic distance.
Human-donor tasks retain the raw HI log$_2$ titres without further residualisation.

\paragraph{GISAID.}
Five variant-era datasets (2021 Wuhan-Hu-1, 2022 Alpha, 2023 BA.1, 2024 XBB.1.5, 2025 KP.3) were constructed from GISAID sequence-count tables aggregated by calendar year.
For each era, every possible single amino-acid substitution or deletion at every Spike position is enumerated relative to the era reference sequence.

\emph{Marginal single-mutation frequency.} For a mutation $x_p\!\to\!y$ (reference residue $x$ at Spike position $p$ replaced by $y$) in era $T$, we define
\begin{align*}
  \mathrm{count}_T(x_p\!\to\!y) &= \text{\# era-$T$ GISAID observations of residue $y$ at position $p$}, \\
  \mathrm{freq}_T(x_p\!\to\!y) &= \frac{\mathrm{count}_T(x_p\!\to\!y)}{\sum_{M}\mathrm{count}_T(M)},
\end{align*}
where the denominator sums over every substitution and deletion tracked at any Spike position in era $T$. ``Marginal'' means marginalising over genetic background: a sequence carrying D614G together with many other mutations contributes 1 to $\mathrm{count}_T(\text{D614G})$, independent of the other mutations it carries. Raw observation counts (\texttt{count\_T}, \texttt{count\_T1}) and global frequencies (\texttt{freq\_T}, \texttt{freq\_T1}) are computed from the preceding- and current-year sequence pools respectively.

\emph{Emergence label.} Letting $\Delta f = \mathrm{freq}_{T+1} - \mathrm{freq}_T$, a mutation is labelled emergent iff
\[
  \texttt{is\_emergent} = 1 \iff \Delta f \ge 10^{-4} \ \ \text{or}\ \ \big(\mathrm{count}_T > 0 \ \text{and}\ \mathrm{freq}_{T+1}/\mathrm{freq}_T \ge 10\big),
\]
the two criteria capturing, respectively, an absolute increase of at least $10^{-4}$ in the era's mutation-observation share and a fold-change criterion inspired by the empirical ``spreading mutation'' threshold (FC~$\ge10$) of \citet{maher2022predicting}. Mutations first observed in the current year are additionally flagged as novel (\texttt{is\_novel}=1).
Model-scoring columns (pLM log-likelihoods) are appended to each era CSV as a post-processing step and are not part of the base reconstruction.

\emph{Threshold sensitivity.} To verify that the model ranking is not an artefact of the threshold choice, we swept the absolute-frequency threshold $F\in\{10^{-3},10^{-4},10^{-5}\}$ and the fold-change threshold $M\in\{5,10,20\}$ and recomputed both the ProGen2-XL metrics (Table~\ref{tab:emergent_sweep}) and the full 25-baseline leaderboard (Table~\ref{tab:emergent_rankpres}), averaged over the five SARS-CoV-2 eras. ProGen2-XL's Top-10\% Recall stays within $[0.359, 0.435]$ and AUROC within $[0.717, 0.771]$ across the grid, and the 25-baseline leaderboard correlates with the paper default ($F=10^{-4}, M=10$) at $\rho\ge0.974$ for every threshold, with ProGen2-XL ranked \#1 throughout.

\begin{table}[H]
  \caption{\textit{is\_emergent} threshold sensitivity for ProGen2-XL (mean over the 5 SARS-CoV-2 eras). $F$: absolute-frequency threshold; $M$: fold-change threshold.}
  \label{tab:emergent_sweep}
  \centering
  \begin{small}
    \begin{tabular}{llcc}
      \toprule
      $F$ & $M$ & Mean Recall & Mean AUROC \\
      \midrule
      $10^{-5}$ & 5  & 0.359 & 0.717 \\
      $10^{-5}$ & 10 & 0.389 & 0.733 \\
      $10^{-5}$ & 20 & 0.413 & 0.744 \\
      $10^{-4}$ & 5  & 0.367 & 0.721 \\
      $10^{-4}$ & 10 & 0.389 (default) & 0.735 (default) \\
      $10^{-4}$ & 20 & 0.425 & 0.755 \\
      $10^{-3}$ & 5  & 0.366 & 0.725 \\
      $10^{-3}$ & 10 & 0.393 & 0.745 \\
      $10^{-3}$ & 20 & 0.435 & 0.771 \\
      \bottomrule
    \end{tabular}
  \end{small}
\end{table}

\begin{table}[H]
  \caption{Leaderboard stability under \textit{is\_emergent} threshold variation. Each row compares the 25-baseline mean-Top-10\%-Recall leaderboard (over 5 eras) under $(F,M)$ against the paper default ($F=10^{-4}, M=10$).}
  \label{tab:emergent_rankpres}
  \centering
  \begin{small}
    \begin{tabular}{llccc}
      \toprule
      $F$ & $M$ & Spearman $\rho$ vs.\ default & top-3 preserved & top-5 preserved \\
      \midrule
      $10^{-5}$ & 5  & 0.974 & 2/3 & 4/5 \\
      $10^{-5}$ & 10 & 0.981 & 2/3 & 5/5 \\
      $10^{-5}$ & 20 & 0.981 & 2/3 & 5/5 \\
      $10^{-4}$ & 5  & 0.987 & 3/3 & 4/5 \\
      $10^{-4}$ & 10 & 1.000 (reference) & 3/3 & 5/5 \\
      $10^{-4}$ & 20 & 0.992 & 2/3 & 5/5 \\
      $10^{-3}$ & 5  & 0.988 & 3/3 & 4/5 \\
      $10^{-3}$ & 10 & 0.995 & 3/3 & 4/5 \\
      $10^{-3}$ & 20 & 0.983 & 3/3 & 4/5 \\
      \bottomrule
    \end{tabular}
  \end{small}
\end{table}

\section{Impact Statement}
\label{sec:impact}
This work highlights a direction for evaluating and applying pLMs: beyond reproducing DMS outcomes, they may capture mutational patterns observed during natural viral evolution, as demonstrated here for SARS-CoV-2. By benchmarking against both experimentally grounded and naturally occurring mutations, our framework suggests that pLMs can provide actionable insights for vaccine design, surveillance, and therapeutic development. These findings support pLMs as complementary tools to experimental assays, with the combined approach offering potential to guide and prioritise future experiments.

We also recognise a potential dual-use risk: tools that accurately predict which mutations are likely to spread could, in principle, be misused to inform the engineering of more transmissible or immune-evasive variants. ViroGym evaluates only existing pre-trained models in a zero-shot setting and does not introduce new generative capabilities; nonetheless, we encourage the community to consider responsible access controls when deploying mutation prioritisation tools in applied settings.

\clearpage
\section*{NeurIPS Paper Checklist}

\begin{enumerate}

\item {\bf Claims}
    \item[] Question: Do the main claims made in the abstract and introduction accurately reflect the paper's contributions and scope?
    \item[] Answer: \answerYes{}
    \item[] Justification: The abstract and introduction state three concrete contributions: a 79-assay DMS benchmark, 21 neutralisation tasks, and a real-world GISAID pandemic prediction task. All claims (autoregressive models outperform masked LMs; DMS/neutralisation performance predicts real-world generalisation; temporal robustness across five variant eras) are directly supported by the experimental results in Sections~3 and~4.

\item {\bf Limitations}
    \item[] Question: Does the paper discuss the limitations of the work performed by the authors?
    \item[] Answer: \answerYes{}
    \item[] Justification: A dedicated ``Limitations and Future Work'' section discusses: (1) the pandemic prediction task is SARS-CoV-2 Spike-specific; (2) cross-viral extension is currently infeasible due to limited GISAID data for other viruses; (3) pseudo-temporal validation does not account for epistatic context or lineage emergence dynamics.

\item {\bf Theory assumptions and proofs}
    \item[] Question: For each theoretical result, does the paper provide the full set of assumptions and a complete (and correct) proof?
    \item[] Answer: \answerNA{}
    \item[] Justification: The paper is an empirical benchmark; it contains no theoretical results, theorems, or proofs.

    \item {\bf Experimental result reproducibility}
    \item[] Question: Does the paper fully disclose all the information needed to reproduce the main experimental results of the paper to the extent that it affects the main claims and/or conclusions of the paper (regardless of whether the code and data are provided or not)?
    \item[] Answer: \answerYes{}
    \item[] Justification: All evaluated models are publicly available pre-trained pLMs used zero-shot (no training required). Dataset sources are listed in Appendix Tables~\ref{DMS_all_reference}--\ref{neutralisation_all_reference}. Scoring strategies are described in Section~3.3 and Appendix~\ref{sec:appendix_results}. Processing pipelines are detailed in Appendix~\ref{sec:appendix_processing}.

\item {\bf Open access to data and code}
    \item[] Question: Does the paper provide open access to the data and code, with sufficient instructions to faithfully reproduce the main experimental results, as described in supplemental material?
    \item[] Answer: \answerYes{}
    \item[] Justification: The processed benchmark datasets and evaluation code are provided as anonymised supplemental material. All underlying DMS and neutralisation assays are drawn from publicly available published sources (cited in Appendix Tables~\ref{DMS_all_reference}--\ref{neutralisation_all_reference}); GISAID data is accessible under the GISAID access agreement.

\item {\bf Experimental setting/details}
    \item[] Question: Does the paper specify all the training and test details (e.g., data splits, hyperparameters, how they were chosen, type of optimizer) necessary to understand the results?
    \item[] Answer: \answerYes{}
    \item[] Justification: All models are evaluated zero-shot with no training or fine-tuning. Scoring strategies for encoder and decoder models are described in Section~3.3 and ablated in Appendix~\ref{sec:appendix_results}. Evaluation metrics (top-10\% recall, absolute Spearman correlation, Precision@K) are defined in Section~3.4. Dataset processing is described in Appendix~\ref{sec:appendix_processing}.

\item {\bf Experiment statistical significance}
    \item[] Question: Does the paper report error bars suitably and correctly defined or other appropriate information about the statistical significance of the experiments?
    \item[] Answer: \answerYes{}
    \item[] Justification: DMS and neutralisation tables report standard deviations across tasks. Pairwise statistical significance for DMS model comparisons is assessed using an exact two-sided paired sign test with Benjamini--Hochberg correction (adjusted $p < 0.05$); full results are in Appendix~\ref{sec:appendix_rankings} Table~\ref{tab:dms_significance}.

\item {\bf Experiments compute resources}
    \item[] Question: For each experiment, does the paper provide sufficient information on the computer resources (type of compute workers, memory, time of execution) needed to reproduce the experiments?
    \item[] Answer: \answerNA{}
    \item[] Justification: All experiments are zero-shot inference on publicly available pre-trained models; no model training was performed. Detailed GPU specifications and inference times are not reported, though all models can be run on standard academic GPU hardware using their publicly released checkpoints.
    
\item {\bf Code of ethics}
    \item[] Question: Does the research conducted in the paper conform, in every respect, with the NeurIPS Code of Ethics \url{https://neurips.cc/public/EthicsGuidelines}?
    \item[] Answer: \answerYes{}
    \item[] Justification: The research uses publicly available experimental data and pre-trained models, involves no human subjects, and does not generate new biological agents. Potential dual-use implications are acknowledged in the Impact Statement.

\item {\bf Broader impacts}
    \item[] Question: Does the paper discuss both potential positive societal impacts and negative societal impacts of the work performed?
    \item[] Answer: \answerYes{}
    \item[] Justification: The Impact Statement (Appendix~\ref{sec:impact}) discusses positive applications including vaccine strain selection, viral surveillance, and therapeutic development. It also acknowledges the dual-use risk that mutation prioritisation tools could in principle inform engineering of more transmissible variants, which motivates responsible access to benchmark data.
    
\item {\bf Safeguards}
    \item[] Question: Does the paper describe safeguards that have been put in place for responsible release of data or models that have a high risk for misuse (e.g., pre-trained language models, image generators, or scraped datasets)?
    \item[] Answer: \answerNA{}
    \item[] Justification: The paper does not release new pre-trained models. Benchmark datasets are curated from published experimental literature; GISAID data is accessed under the GISAID access agreement, which itself provides a usage governance framework.

\item {\bf Licenses for existing assets}
    \item[] Question: Are the creators or original owners of assets (e.g., code, data, models), used in the paper, properly credited and are the license and terms of use explicitly mentioned and properly respected?
    \item[] Answer: \answerYes{}
    \item[] Justification: All DMS and neutralisation datasets are cited with their original publications (Appendix Tables~\ref{DMS_all_reference}--\ref{neutralisation_all_reference}). All pre-trained pLMs are cited and used in accordance with their publicly released terms. GISAID data is used under the GISAID access agreement.

\item {\bf New assets}
    \item[] Question: Are new assets introduced in the paper well documented and is the documentation provided alongside the assets?
    \item[] Answer: \answerYes{}
    \item[] Justification: The ViroGym benchmark (processed DMS assays, neutralisation tasks, GISAID evaluation data, and evaluation code) is documented in the main paper and appendix. Dataset provenance, processing pipelines, and schema are described in Sections~3.1--3.2 and Appendix~\ref{sec:appendix_processing}. Anonymised assets are provided as supplemental material.

\item {\bf Crowdsourcing and research with human subjects}
    \item[] Question: For crowdsourcing experiments and research with human subjects, does the paper include the full text of instructions given to participants and screenshots, if applicable, as well as details about compensation (if any)?
    \item[] Answer: \answerNA{}
    \item[] Justification: The paper does not involve crowdsourcing or research with human subjects. All data is sourced from published experimental assays and genomic surveillance databases.

\item {\bf Institutional review board (IRB) approvals or equivalent for research with human subjects}
    \item[] Question: Does the paper describe potential risks incurred by study participants, whether such risks were disclosed to the subjects, and whether Institutional Review Board (IRB) approvals (or an equivalent approval/review based on the requirements of your country or institution) were obtained?
    \item[] Answer: \answerNA{}
    \item[] Justification: The paper does not involve human subjects research; all data is from published experimental assays and genomic surveillance databases. No IRB approval is required.

\item {\bf Declaration of LLM usage}
    \item[] Question: Does the paper describe the usage of LLMs if it is an important, original, or non-standard component of the core methods in this research? Note that if the LLM is used only for writing, editing, or formatting purposes and does \emph{not} impact the core methodology, scientific rigor, or originality of the research, declaration is not required.
    \item[] Answer: \answerNA{}
    \item[] Justification: The protein language models evaluated in this paper are the \emph{subject} of benchmarking, not novel methodological components introduced by the authors. No LLMs were used as important or non-standard components of the core research methodology.

\end{enumerate}

\end{document}